\documentclass{amsart}
\usepackage{hyperref}
\usepackage{graphicx}
\usepackage{curves}

\newcommand*{\mailto}[1]{\href{mailto:#1}{\nolinkurl{#1}}}

\newtheorem{theorem}{Theorem}[section]
\newtheorem{lemma}[theorem]{Lemma}
\newtheorem{corollary}[theorem]{Corollary}
\newtheorem{remark}[theorem]{Remark}
\newtheorem{hypothesis}[theorem]{Hypothesis {\bf H.}\hspace*{-0.6ex}}

\newcommand{\R}{{\mathbb R}}
\newcommand{\N}{{\mathbb N}}
\newcommand{\Z}{{\mathbb Z}}
\newcommand{\C}{{\mathbb C}}
\newcommand{\M}{{\mathbb M}}

\newcommand{\nn}{\nonumber}
\newcommand{\be}{\begin{equation}}
\newcommand{\ee}{\end{equation}}
\newcommand{\bea}{\begin{eqnarray}}
\newcommand{\eea}{\end{eqnarray}}
\newcommand{\ul}{\underline}
\newcommand{\ol}{\overline}
\newcommand{\ti}{\tilde}

\newcommand{\spr}[2]{\langle #1 , #2 \rangle}
\newcommand{\id}{\mathbb{I}}
\newcommand{\I}{\mathrm{i}}
\newcommand{\E}{\mathrm{e}}
\newcommand{\ind}{\mathrm{ind}}
\newcommand{\re}{\mathrm{Re}}
\newcommand{\im}{\mathrm{Im}}
\newcommand{\floor}[1]{\lfloor#1 \rfloor}
\def\Xint#1{\mathchoice
   {\XXint\displaystyle\textstyle{#1}}%
   {\XXint\textstyle\scriptstyle{#1}}%
   {\XXint\scriptstyle\scriptscriptstyle{#1}}%
   {\XXint\scriptscriptstyle\scriptscriptstyle{#1}}%
   \!\int}
\def\XXint#1#2#3{{\setbox0=\hbox{$#1{#2#3}{\int}$}
     \vcenter{\hbox{$#2#3$}}\kern-.5\wd0}}
\def\dashint{\Xint-}
\newcommand{\sigI}{\begin{pmatrix} 0 & 1 \\ 1 & 0 \end{pmatrix}}
\newcommand{\rI}{\begin{pmatrix}  1 & 1 \end{pmatrix}}

\newcommand{\ulz}{\underline{z}}
\newcommand{\di}{\mathcal{D}}
\newcommand{\vrc}{\ul{\Xi}_{p_0}}
\newcommand{\hvrc}{\ul{\hat{\Xi}}_{p_0}}
\newcommand{\hmu}{\hat{\mu}}

\newcommand{\uhmuz}{{\underline{\hat{\mu}}}}
\newcommand{\hnu}{\hat{\nu}}
\newcommand{\uhnu}{{\underline{\hat{\nu}}(n,t)}}
\newcommand{\uhnuz}{{\underline{\hat{\nu}}}}
\newcommand{\dimu}[1]{\di_{\ul{\hat{\mu}}(#1)}}
\newcommand{\dimus}[1]{\di_{\ul{\hat{\mu}}(#1)^*}}
\newcommand{\dimuz}{\di_{\ul{\hat{\mu}}}}
\newcommand{\dimuzs}{\di_{\ul{\hat{\mu}}^*}}
\newcommand{\dinu}[1]{\di_{\ul{\hat{\nu}}(#1)}}
\newcommand{\dinus}[1]{\di_{\ul{\hat{\nu}}(#1)^*}}
\newcommand{\dinuz}{\di_{\ul{\hat{\nu}}}}

\newcommand{\Amap}{\ul{A}_{p_0}}
\newcommand{\amap}{\ul{\alpha}_{p_0}}
\newcommand{\hAmap}{\ul{\hat{A}}_{p_0}}
\newcommand{\hamap}{\ul{\hat{\alpha}}_{p_0}}

\newcommand{\Rg}[1]{R_{2g+2}^{1/2}(#1)}

\newcommand{\eps}{\varepsilon}

\newcommand{\sig}{\sigma}
\newcommand{\lam}{\lambda}

\newcommand{\om}{\omega}

\numberwithin{equation}{section}

\begin{document}

\title[Long-Time Asymptotics of the Perturbed Periodic Toda Lattice]{Long-Time Asymptotics of the Periodic Toda Lattice under Short-Range Perturbations}

\author[S. Kamvissis]{Spyridon Kamvissis}
\address{Department of Applied Mathematics, University of Crete \\
714 09 Knossos, Greece}
\email{\mailto{spyros@tem.uoc.gr}}
\urladdr{\url{http://www.tem.uoc.gr/~spyros/}}

\author[G. Teschl]{Gerald Teschl}
\address{Faculty of Mathematics\\ University of Vienna\\
Nordbergstrasse 15\\ 1090 Wien\\ Austria\\ and International Erwin Schr\"odinger
Institute for Mathematical Physics\\ Boltzmanngasse 9\\ 1090 Wien\\ Austria}
\email{\mailto{Gerald.Teschl@univie.ac.at}}
\urladdr{\url{http://www.mat.univie.ac.at/~gerald/}}

\thanks{J. Math. Phys. {\bf 53}, 073706 (2012)}
\thanks{Research supported in part by the ESF programme MISGAM, and
the Austrian Science Fund (FWF) under Grant No.\ P17762 and Y330.}


\begin{abstract}
We consider the long-time asymptotics of periodic (and slightly more generally of algebro-geometric
finite-gap) solutions of the doubly infinite Toda lattice 
under a short-range perturbation. We prove that the perturbed lattice asymptotically
approaches a modulated lattice.

More precisely, let $g$ be the genus of the hyperelliptic curve associated with
the unperturbed solution. We show that, apart from the phenomenon of
solitons travelling on the quasi-periodic background, the $n/t$-pane 
contains  $g+2$ areas where
the perturbed solution is close to a finite-gap solution on the same isospectral
torus. In between there are $g+1$ regions where the perturbed solution is asymptotically
close to a modulated lattice which undergoes a
continuous phase transition (in the Jacobian variety) and which interpolates
between these isospectral solutions. In the special case of  the free lattice ($g=0$) the
isospectral torus consists of just one point and we recover the known result.

Both the solutions in the isospectral torus and the phase transition are explicitly
characterized in terms of Abelian integrals on the underlying hyperelliptic curve.

Our method relies on the equivalence of the inverse spectral problem to
a vector Riemann--Hilbert problem defined on the hyperelliptic curve
and generalizes the so-called nonlinear 
stationary phase/steepest descent method for
Riemann--Hilbert problem deformations to Riemann surfaces.
\end{abstract}

\maketitle

\section{Introduction}

A classical result going back to
Zabusky and Kruskal \cite{zakr} states that a decaying (fast enough) perturbation
of the constant solution of a soliton equation eventually splits into a number of   "solitons":
localized travelling waves that preserve their shape and velocity after interaction, plus a decaying radiation part.
This is the  motivation
for the result presented here. Our aim is 
to investigate the case where the constant background
solution is replaced by a periodic one. 
We provide the detailed analysis in the case of the Toda lattice though it is
clear that our methods apply to other soliton equations as well.

In the case of the Korteweg--de Vries equation the asymptotic result
was first shown  by \v{S}abat \cite{sh} and by Tanaka \cite{ta2}.
Precise asymptotics for the radiation part were first formally derived by Zakharov and
Manakov \cite{zama} and by Ablowitz and Segur \cite{as}, \cite{as2} with further extensions by
Buslaev and Sukhanov \cite{bs}. A detailed rigorous justification not requiring any a priori
information on the asymptotic form of the solution was first given by Deift and Zhou \cite{dz}
for the case of the modified Korteweg--de Vries equation,
inspired by earlier work of Manakov \cite{ma} and Its \cite{its1} (see also \cite{its2}, \cite{its3}, \cite{ip}).
For further information on the history of this problem we refer to the survey by
Deift, Its, and Zhou \cite{diz}.

A naive guess would be that the perturbed periodic lattice approaches the
unperturbed one in the uniform norm. However,
as pointed out in \cite{kt} this is wrong:
\begin{figure}
\includegraphics[width=8cm]{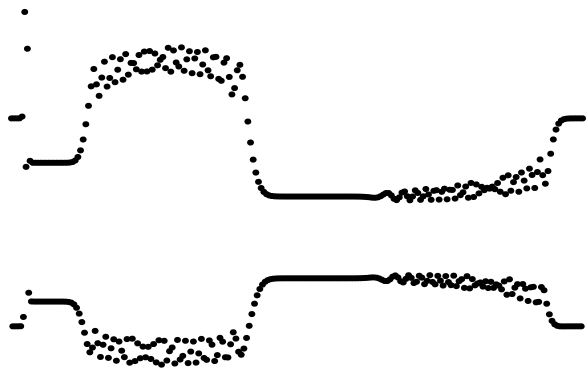}
\caption{Numerically computed solution of the Toda lattice, with initial
condition a period two solution perturbed at one point in the middle.} \label{fig1}
\end{figure}
In Figure~\ref{fig1} the two observed lines express the variables $a(n,t)$ 
of the Toda lattice (see \eqref{TLpert} below) at a frozen time $t$.
In areas where the lines seem to be continuous this is due to the fact that we have plotted
a huge number of particles and also due to the 2-periodicity in space. So one can think of
the two lines as the even- and odd-numbered particles of the lattice.
We first note  the single soliton which separates two regions of
apparent periodicity on the left. Also, after the soliton, we observe three different 
areas with apparently periodic solutions of  period two.
Finally there are some transitional regions in between
which interpolate between the different period two regions. It is the
purpose of this paper to give a rigorous and complete mathematical explanation of
this picture. This will be done by formulating the inverse spectral problem as
a vector Riemann--Hilbert problem on the underlying hyperelliptic curve and
extending the nonlinear steepest descent method to this new setting.
While Riemann--Hilbert problem on Riemann surfaces have been considered in
detail before, see for example the monograph by Rodin \cite{ro}, we extend this theory
as well (see e.g.\ our novel solution formula for scalar Riemann--Hilbert problems in
Theorem~\ref{thmplf}).

Consider the doubly infinite Toda lattice in Flaschka's
variables (see e.g.\ \cite{ft}, \cite{tjac}, \cite{taet}, or \cite{ta})
\be \label{TLpert}
\aligned
\dot b(n,t) &= 2(a(n,t)^2 -a(n-1,t)^2),\\
\dot a(n,t) &= a(n,t) (b(n+1,t) -b(n,t)),
\endaligned
\ee
$(n,t) \in \Z \times \R$,
where the dot denotes differentiation with respect to time.

In case of a constant background the long-time asymptotics
were first computed by Novokshenov and Habibullin \cite{nh} and later made
rigorous by Kamvissis \cite{km} under  the additional assumption that no solitons
are present. The full case (with solitons) was only recently 
presented  by Kr\"uger and Teschl in \cite{krt}
(for a review see also \cite{krt2}). 

Here we will consider a quasi-periodic algebro-geometric background
solution $(a_q,b_q)$, to be described in the next section, plus
a short-range perturbation $(a,b)$ satisfying
\be \label{decay}
\sum_{n\in\Z} n^6 (|a(n,t) - a_q(n,t)| + |b(n,t)- b_q(n,t)|) < \infty
\ee
for $t=0$  and hence for all  (see e.g. \cite{emt2}) $t\in\R$. The
perturbed solution can be computed via the inverse
scattering transform. The case where $(a_q,b_q)$ is constant
is classical (see again \cite{ft}, \cite{tjac} or \cite{ta}), 
while the more general case we want here was solved only recently in \cite{emt2}
(see also \cite{mt}).

To fix our background solution, consider a hyperelliptic Riemann surface of genus $g$
with real moduli $E_0, E_1, ...., E_{2g+1}$. Choose a Dirichlet divisor $\dimuz$ and introduce
\be
\ulz(n,t) = \hAmap(\infty_+) - \hamap(\dimuz) - n\ul{\hat A}_{\infty_-}(\infty_+)
+ t\ul{U}_0 - \hvrc \in \C^g,
\ee
where $\Amap$ ($\amap$) is Abel's map (for divisors) and $\vrc$, $\ul{U}_0$ are 
some constants defined in Section~\ref{secAG}. Then our background solution is given in terms
of Riemann theta functions (defined in (2.14)) by
\begin{align} \nn
a_q(n,t)^2 &= \ti{a}^2 \frac{\theta(\ulz(n+1,t)) \theta(\ulz(n-1,t))}{\theta(
\ulz(n,t))^2},\\
b_q(n,t) &= \tilde{b} + \frac{1}{2}
\frac{d}{dt} \log\Big(\frac{\theta(\ulz(n,t)) }{\theta(\ulz(n-1,t))}\Big),
\end{align}
where $\ti{a}$, $\tilde{b} \in\R$ are again some constants.

We can of course view this hyperelliptic Riemann surface as formed by cutting and pasting
two copies of the complex plane along bands. Having this picture
in mind, we denote the standard projection  
to the complex plane by $\pi$. 

Assume for simplicity that the Jacobi operator
\be
H(t) f(n) = a(n,t) f(n+1) + a(n-1,t) f(n-1) + b(n,t) f(n), \quad f\in\ell^2(\Z),
\ee
corresponding to the perturbed problem \eqref{TLpert} has no eigenvalues.
In this paper we prove that for long times the perturbed Toda lattice
is  asymptotically close to the following limiting lattice 
defined by
\be \label{limlat}
\aligned
\prod_{j=n}^{\infty} (\frac{a_l(j,t)}{a_q(j,t)} )^2 = &
\frac{\theta(\ulz(n,t))}{\theta(\ulz(n-1,t))}
\frac{\theta(\ulz(n-1,t)+ \ul{\delta}(n,t))}{\theta(\ulz(n,t)+\ul{\delta}(n,t))} \times\\
& \times \exp\left( \frac{1}{2\pi\I} \int_{C(n/t)} 
\log (1-|R|^2) \om_{\infty_+\, \infty_-}\right),\\
\delta_\ell(n,t)= & \frac{1}{2\pi\I} \int_{C(n/t)} \log (1-|R|^2) \zeta_\ell,
\endaligned
\ee
where $R$ is the associated reflection coefficient, $\zeta_\ell$ is a canonical basis
of holomorphic differentials, $\om_{\infty_+\, \infty_-}$ is an Abelian differential of
the third kind defined in \eqref{ominfpm}, and $C(n/t)$ is a contour on the Riemann
surface. More specific, $C(n/t)$ is obtained by taking the spectrum of the unperturbed
Jacobi operator $H_q$ between $-\infty$ and a special stationary phase point $z_j(n/t)$,
for the phase of the underlying Riemann--Hilbert problem defined in
the beginning of Section~\ref{secSPP}, and lifting it to the Riemann surface
(oriented such that the upper sheet lies to its left). The point $z_j(n/t)$ will move from
$-\infty$ to $+\infty$ as $n/t$ varies from $-\infty$ to $+\infty$.
From the products above, one easily recovers $a_l(n,t)$.
More precisely, we have the following.

\begin{theorem}\label{thmMain}
Let $C$ be any  (large) positive number and $\delta$ be any (small)
positive number. Let $E_s \in S$ be the 'resonance points' defined by
$S= \{ E_s : |R(E_s)| =1\}.$ (There are at most $2g+2$ such points, since they are always
endpoints $E_j$ of the bands that constitute the spectrum of the Jacobi operator.) 
Consider the  region 
$D= \{(n,t): |\frac{n}{t}| < C \} \cap \{(n,t): |z_j(\frac{n}{t})-E_s|>\delta$\}, 
where  $z_j(\frac{n}{t})$ is the special stationary phase point for the phase defined in
the beginning of Section~\ref{secSPP}. Then one has
\be
\prod_{j=n}^{\infty} \frac{a_l(j,t)}{a(j,t)} \to 1
\ee
uniformly in $D$, as $t \to \infty$.
\end{theorem}

The proof of this theorem will be given in Section~\ref{secSPP} of this paper.

\begin{remark}
(i) It is easy to see how the asymptotic formula above describes the picture
given by the numerics. Recall that the spectrum $\sig(H_q)$ of $H_q$ consists
of $g+1$ bands whose band edges are the branch points of the underlying
hyperelliptic Riemann surface. If $\frac{n}{t}$ is small enough, $z_j(n/t)$ is to the
left of all bands implying that $C(n/t)$ is empty and thus $\delta_\ell(n,t)=0$;
so we recover the purely periodic lattice.
At some value of  $\frac{n}{t}$  a stationary phase point first appears in the
first band of $\sig(H_q)$ and begins to move form
the left endpoint of the band towards the right endpoint of the band.
(More precisely we have a pair of stationary phase points $z_j$ and $z_j^*$, one in each sheet
of the hyperelliptic curve, with common projection $\pi(z_j)$ on the complex plane.)
So $\delta_\ell(n,t)$ is now a non-zero quantity changing with $\frac{n}{t}$
and the asymptotic lattice has a slowly modulated non-zero phase. 
Also the factor given by the exponential of the integral is 
non-trivially changing with $\frac{n}{t}$ and contributes  to a
slowly modulated amplitude. Then, after the
stationary phase point leaves the first band
there is a range of $\frac{n}{t}$ for which  
no stationary phase point appears in the spectrum $\sig(H_q)$, hence the phase shift
$\delta_\ell(n,t)$ and the integral remain constant, so the asymptotic lattice is periodic
(but with a non-zero phase shift). Eventually a stationary phase point
appears in the second band, so a new modulation appears and so on.
Finally, when $\frac{n}{t}$ is large enough, so that all bands have been
traversed by the stationary phase point(s), the asymptotic lattice is
again periodic. Periodicity properties of theta functions 
easily show that phase shift  is actually cancelled by the exponential
of the integral and we recover the original periodic lattice with no
phase shift at all.

(ii) If eigenvalues are present we can apply appropriate Darboux transformations
to add the effect of such eigenvalues (\cite{emt4}). What we then see asymptotically is
travelling solitons in a periodic background. Note that this will change the
asymptotics on one side. In any case, our method works unaltered for such 
situations (cf.\ \cite{emt3}) as well.

(iii) Employing the very same methods of the paper it is 
very easy to show that in any region
$|\frac{n}{t}|> C$, one has
\be
\prod_{j=n}^{\infty} \frac{a_l(j,t)}{a(j,t)} \to 1
\ee
uniformly in $t$, as $n \to \infty$.

(iv) The effect of the resonances $E_s$ is only felt locally (and to higher 
order in $1/t$) in some small
(decaying as $t \to \infty$) region, where in fact $ |z_j(\frac{n}{t})-E_s| \to 0$
as $t \to \infty$. So the above theorem is actually true in 
$\{(n,t): |\frac{n}{t}| < C \}$.
Near the resonances we expect both a "collisionless shock" phenomenon and
a Painlev\'e region
to appear (\cite{dvz}, \cite{dz}, \cite{km}, \cite{km2}). A proof of this can be given using the results
of \cite{dvz} and \cite{dz}.

(v) For the proof of Theorem~\ref{thmMain} and  Theorem~\ref{thmMain2} it
would suffice to assume \eqref{decay} with $n^6$ replaced by $|n|^3$ (or even
$|n|$ plus the requirement that the associated reflection coefficient is H\"older continuous).
Our stronger assumption is only required for the detailed decay estimates in
Theorem~\ref{thmMain3} below.
\end{remark}

By dividing in \eqref{limlat} one recovers the $a(n,t)$.
It follows from the main Theorem and the last remark above that
\be
|a(n,t) - a_l(n,t)| \to 0
\ee
uniformly in $D$, as $t \to \infty$.
In other words, 
the perturbed  Toda lattice
is  asymptotically close to the  limiting  lattice above.

A similar theorem can be proved for the  velocities $b(n,t)$.

\begin{theorem}\label{thmMain2}
In the  region
$D= \{(n,t): |\frac{n}{t}| < C \} \cap \{(n,t): |z_j(\frac{n}{t})-E_s|>\delta$\},
of Theorem ~\ref{thmMain} we also have
\be
\sum_{j=n}^\infty \big( b_l(j,t) - b_q(j,t) \big) \to 0
\ee
uniformly in $D$, as $t \to \infty$, where $b_l$ is given by
\be
\aligned
\sum_{j=n}^\infty \big( b_l(j,t) - b_q(j,t) \big) = &
\frac{1}{2\pi\I} \int_{C(n/t)} \log(1-|R|^2) \Omega_0\\
& {} + \frac{1}{2}\frac{d}{ds}
\log\left( \frac{\theta(\ulz(n,s) + \ul{\delta}(n,t) )}{\theta(\ulz(n,s))} \right) \Big|_{s=t}
\endaligned
\ee
and $\Omega_0$ is an Abelian differential of the second kind defined in \eqref{Om0}.
\end{theorem}

The proof of this theorem will also be given in Section~\ref{secSPP} of this paper.

The next question we address here concerns the higher order asymptotics.
Namely, what is the rate at which the perturbed lattice approaches the limiting
lattice? Even more, what is the exact asymptotic formula?

\begin{theorem}\label{thmMain3} 
Let $D_j$ be the sector $D_j = \{ (n,t) ,:\, z_j(n/t)\in[E_{2j}+\eps, E_{2j+1}-\eps]$ for some $\eps>0$.
Then one has
\be
\prod_{j=n}^{\infty} \left(\frac{a(j,t)}{a_l(j,t)}\right)^2  = 1 +
\sqrt{\frac{\I}{\phi''(z_j(n/t))t}} 2 \re\left( \ol{\beta(n,t)} \I \Lambda_0(n,t) \right) +
O(t^{-\alpha})
\ee
and
\be
\sum_{j=n+1}^{\infty} \big(b(j,t) - b_l(j,t) \big) = \sqrt{\frac{\I}{\phi''(z_j(n/t))t}} 2 \re\left( \ol{\beta(n,t)} \I\Lambda_1(n,t)\right) +
O(t^{-\alpha})
\ee
for any $\alpha<1$ uniformly in $D_j$, as $t \to \infty$. 
Here 
\be
\phi''(z_j) / \I = \frac{\prod_{k=0,k\ne j}^g (z_j -z_k)}{\I\Rg{z_j}} >0,
\ee
(where $\phi(p,n/t)$ is the phase function defined in \eqref{defsp} and $\Rg{z}$ the square root of the underlying Riemann surface),
\begin{align}\nn
\Lambda_0(n,t) & =
\om_{\infty_-\, \infty_+}(z_j) + \sum_{k,\ell} c_{k\ell}(\uhnu) \int_{\infty_+}^{\infty_-} \om_{\hnu_\ell(n,t),0} \zeta_k(z_j),\\
\Lambda_1(n,t) &=  \om_{\infty_-,0}(z_j) -  \sum_{k,\ell} c_{k\ell}(\uhnu) \om_{\hnu_\ell(n,t),0}(\infty_+) \zeta_k(z_j),
\end{align}
with $c_{k\ell}(\uhnu)$ some constants defined in \eqref{defcjlnu}, $\om_{q,0}$ an Abelian differential of the second kind with
a second order pole at $q$ (cf.\ Remark~\ref{remabdiff2k}),
\begin{align}\nn
\beta =&  \sqrt{\nu} \E^{\I(\pi/4-\arg(R(z_j)))+\arg(\Gamma(\I\nu))-2 \nu \alpha(z_j))}
\left(\frac{\phi''(z_j)}{\I}\right)^{\I\nu}  \E^{-t \phi(z_j)} t^{-\I\nu} \times\\ \nn
& \times \frac{\theta(\ulz(z_j,n,t)+ \ul{\delta}(n,t))}{\theta(\ulz(z_j,0,0))}
\frac{\theta(\ulz(z_j^*,0,0))}{\theta(\ulz(z_j^*,n,t)+ \ul{\delta}(n,t))} \times\\
& \times \exp\left( \frac{1}{2\pi\I} \int_{C(n/t)} \log\left(\frac{1-|R|^2}{1-|R(z_j)|^2}\right) \om_{p\, p^*}\right),
\end{align}
where $\Gamma(z)$ is the gamma function,
\be
\nu = - \frac{1}{2\pi} \log (1-|R(z_j)|^2)>0,
\ee
and $\alpha(z_j)$ is a constant defined in \eqref{def:alzj}.
\end{theorem}

The proof of this theorem will be given in Section~\ref{secCROSS} of this paper.
The idea of the proof is that even when a Riemann-Hilbert problem needs to be
considered on an algebraic variety, a localized parametrix Riemann-Hilbert problem 
need only be solved in the complex plane and the local solution can then be glued
to the global Riemann-Hilbert solution on the variety.

The same idea can produce the asymptotics in the two resonance regions mentioned above:
a "collisionless shock" phenomenon and a Painlev\'e region,
for every resonance pint $E_s$, by simply using the results of 
(\cite{dvz}, \cite{dz}). We leave the details to the reader.

\begin{remark}
(i) The current work combines two articles that have appeared previously in the arXiv as 
arXiv:0705.0346 and arXiv:0805.3847 but have not been published otherwise.
The necessary changes needed to include solitons are given in \cite{krt3} which was based on
arXiv:0705.0346
(see also \cite{emt4}, \cite{krt}, and \cite{tag}).

(ii) Combining our technique with the one from \cite{dz2} can lead to a
complete asymptotic expansion.

(iii) Finally, we note that the same proof works even if there are different spatial
asymptotics as $n\to\pm\infty$ as long as they lie in the same isospectral class (cf.\ \cite{emt3}).
\end{remark}

\section{Algebro-geometric quasi-periodic finite-gap solutions}
\label{secAG}

As a preparation we need some facts on our background
solution $(a_q,b_q)$ which we want to choose from the class of
algebro-geometric quasi-periodic finite-gap solutions, that is the
class of stationary solutions of the Toda hierarchy, \cite{bght}, \cite{ghmt}, \cite{tjac}.
In particular, this class contains all periodic solutions. We will
use the same notation as in \cite{tjac}, where we also refer to for proofs.
As a reference for Riemann surfaces in this context we recommend \cite{fk}.

To set the stage let $\M$ be the Riemann surface associated with the following function
\begin{equation}
\Rg{z}, \qquad R_{2g+2}(z) = \prod_{j=0}^{2g+1} (z-E_j), \qquad
E_0 < E_1 < \cdots < E_{2g+1},
\end{equation}
$g\in \N$. $\M$ is a compact, hyperelliptic Riemann surface of genus $g$.
We will choose $\Rg{z}$ as the fixed branch
\begin{equation}
\Rg{z} = -\prod_{j=0}^{2g+1} \sqrt{z-E_j},
\end{equation}
where $\sqrt{.}$ is the standard root with branch cut along $(-\infty,0)$.

A point on $\M$ is denoted by 
$p = (z, \pm \Rg{z}) = (z, \pm)$, $z \in \C$, or $p = (\infty,\pm) = \infty_\pm$, and
the projection onto $\C \cup \{\infty\}$ by $\pi(p) = z$. 
The points $\{(E_{j}, 0), 0 \leq j \leq 2 g+1\} \subseteq \M$ are 
called branch points and the sets 
\begin{equation}
\Pi_{\pm} = \{ (z, \pm \Rg{z}) \mid z \in \C\setminus
\bigcup_{j=0}^g[E_{2j}, E_{2j+1}]\} \subset \M
\end{equation}
are called upper, lower sheet, respectively.

Let $\{a_j, b_j\}_{j=1}^g$ be loops on the surface $\M$ representing the
canonical generators of the fundamental group $\pi_1(\M)$. We require
$a_j$ to surround the points $E_{2j-1}$, $E_{2j}$ (thereby changing sheets
twice) and $b_j$ to surround $E_0$, $E_{2j-1}$ counterclockwise on the
upper sheet, with pairwise intersection indices given by
\begin{equation}
a_i \circ a_j= b_i \circ b_j = 0, \qquad a_i \circ b_j = \delta_{i,j},
\qquad 1 \leq i, j \leq g.
\end{equation}
The corresponding canonical basis $\{\zeta_j\}_{j=1}^g$ for the space of
holomorphic differentials can be constructed by
\begin{equation}
\underline{\zeta} = \sum_{j=1}^g \underline{c}(j)  
\frac{\pi^{j-1}d\pi}{R_{2g+2}^{1/2}},
\end{equation}
where the constants $\underline{c}(.)$ are given by
\be\label{defcjk}
c_j(k) = C_{jk}^{-1}, \qquad 
C_{jk} = \int_{a_k} \frac{\pi^{j-1}d\pi}{R_{2g+2}^{1/2}} =
2 \int_{E_{2k-1}}^{E_{2k}} \frac{z^{j-1}dz}{\Rg{z}} \in
\R.
\ee
The differentials fulfill
\begin{equation} \label{deftau}
\int_{a_j} \zeta_k = \delta_{j,k}, \qquad \int_{b_j} \zeta_k = \tau_{j,k}, 
\qquad \tau_{j,k} = \tau_{k, j}, \qquad 1 \leq j, k \leq g.
\end{equation}

Now pick $g$ numbers (the Dirichlet eigenvalues)
\be
(\hat{\mu}_j)_{j=1}^g = (\mu_j, \sigma_j)_{j=1}^g
\ee
whose projections lie in the spectral gaps, that is, $\mu_j\in[E_{2j-1},E_{2j}]$.
Associated with these numbers is the divisor $\dimuz$ which
is one at the points $\hat{\mu}_j$  and zero else. Using this divisor we
introduce
\begin{align} \nn
\ulz(p,n,t) &= \hAmap(p) - \hamap(\dimuz) - n\ul{\hat A}_{\infty_-}(\infty_+)
+ t\ul{U}_0 - \hvrc \in \C^g, \\
\ulz(n,t) &= \ulz(\infty_+,n,t),
\end{align}
where $\vrc$ is the vector of Riemann constants
\begin{equation}
\hat{\Xi}_{p_0,j} = \frac{j+ \sum_{k=1}^g \tau_{j,k}}{2},
\qquad p_0=(E_0,0),
\end{equation}
$\ul{U}_0$ are the $b$-periods of the Abelian differential $\Omega_0$ defined below,
and $\Amap$ ($\amap$) is Abel's map (for divisors). The hat indicates that we
regard it as a (single-valued) map from $\hat{\M}$ (the fundamental polygon
associated with $\M$ by cutting along the $a$ and $b$ cycles) to $\C^g$.
We recall that the function $\theta(\ulz(p,n,t))$ has precisely $g$ zeros
$\hmu_j(n,t)$ (with $\hmu_j(0,0)=\hmu_j$), where $\theta(\ul{z})$ is the
Riemann theta function of $\M$.

Then our background solution is given by
\begin{align} \nn
a_q(n,t)^2 &= \ti{a}^2 \frac{\theta(\ulz(n+1,t)) \theta(\ulz(n-1,t))}{\theta(
\ulz(n,t))^2},\\ \label{imfab}
b_q(n,t) &= \tilde{b} + \frac{1}{2}
\frac{d}{dt} \log\Big(\frac{\theta(\ulz(n,t)) }{\theta(\ulz(n-1,t))}\Big).
\end{align}
The constants $\ti{a}$, $\tilde{b}$ depend only on the Riemann surface
(see \cite[Section~9.2]{tjac}).

Introduce the time dependent Baker-Akhiezer function
\begin{align}\label{defpsiq}
\psi_q(p,n,t) &= C(n,0,t) \frac{\theta (\ulz(p,n,t))}{\theta(\ulz (p,0,0))}
\exp \Big( n \int_{E_0}^p \om_{\infty_+\, \infty_-} + t\int_{E_0}^p \Omega_0
\Big),
\end{align}
where $C(n,0,t)$ is real-valued,
\begin{equation}
C(n,0,t)^2 = \frac{ \theta(\ulz(0,0)) \theta(\ulz(-1,0))}
{\theta (\ulz (n,t))\theta (\ulz (n-1,t))},
\end{equation}
and the sign has to be chosen in accordance with $a_q(n,t)$.
Here
\be
\theta(\ulz) = \sum_{\ul{m} \in \Z^g} \exp 2 \pi \I \left( \spr{\ul{m}}{\ulz} +
\frac{\spr{\ul{m}}{\ul{\tau} \, \ul{m}}}{2}\right) ,\qquad \ulz \in \C^g,
\ee
is the Riemann theta function associated with $\M$,
\be\label{ominfpm}
\om_{\infty_+\, \infty_-}= \frac{\prod_{j=1}^g (\pi -\lambda_j) }{R_{2g+2}^{1/2}}d\pi
\ee
is the Abelian differential of the third kind with poles at $\infty_+$ and $\infty_-$ and
\be\label{Om0}
\Omega_0 = \frac{\prod_{j=0}^g (\pi - \ti\lambda_j) }{R_{2g+2}^{1/2}}d\pi,
\qquad \sum_{j=0}^g \ti\lambda_j = \frac{1}{2} \sum_{j=0}^{2g+1} E_j,
\ee
is the Abelian differential of the second kind with second order poles at
$\infty_+$ respectively $\infty_-$ (see \cite[Sects.~13.1, 13.2]{tjac}).
All Abelian differentials are normalized to have vanishing $a_j$ periods.

The Baker-Akhiezer function is a meromorphic function on $\M\setminus\{\infty_\pm\}$
with an essential singularity at $\infty_\pm$. The two branches are denoted by
\begin{equation}
\psi_{q,\pm}(z,n,t) = \psi_q(p,n,t), \qquad p=(z,\pm)
\end{equation}
and it satisfies
\begin{align}\nn
H_q(t) \psi_q(p,n,t) &= \pi(p) \psi_q(p,n,t),\\
\frac{d}{dt} \psi_q(p,n,t) &= P_{q,2}(t) \psi_q(p,n,t),
\end{align}
where
\begin{align}
H_q(t) f(n) &= a_q(n,t) f(n+1) + a_q(n-1,t) f(n-1) + b_q(n,t) f(n),\\
P_{q,2}(t) f(n) &= a_q(n,t) f(n+1) - a_q(n-1,t) f(n-1)
\end{align}
are the operators from the Lax pair for the Toda lattice.

It is well known that the spectrum of $H_q(t)$ is time independent and
consists of $g+1$ bands
\begin{equation}
\sig(H_q) = \bigcup_{j=0}^g [E_{2j},E_{2j+1}].
\end{equation}
For further information and proofs we refer to \cite[Chap.~9 and Sect.~13.2]{tjac}.

\section{The Inverse scattering transform and the Riemann--Hilbert problem}
\label{secISTRH}

In this section our notation and results are  taken from \cite{emt} and \cite{emt2}.
Let $\psi_{q,\pm}(z,n,t)$ be the branches of the Baker-Akhiezer function defined
in the previous section. Let $\psi_\pm(z,n,t)$ be the Jost functions for the perturbed
problem 
\be
a(n,t) \psi_\pm(z,n+1,t) + a(n-1,t) \psi_\pm(z,n-1,t) + b(n,t) \psi_\pm(z,n,t) =z  \psi_\pm(z,n,t)
\ee
defined by the asymptotic normalization
\be
\lim_{n \to \pm \infty} 
w(z)^{\mp  n} ( \psi_\pm(z,n,t) - \psi_{q, \pm}(z,n,t))
=0,
\ee
where $w(z)$ is the quasimomentum map
\be
w(z)= \exp (\int^p_{E_0} \om_{\infty_+\, \infty_-}), \quad p=(z,+).
\ee
The asymptotics of the  two projections of the Jost function are
\begin{align} \nn
\psi_\pm(z,n,t) =& \, \psi_{q,\pm}(z,0,t)
\frac{z^{\mp n} \Big(\prod_{j=0}^{n-1} a_q(j,t)\Big)^{\pm 1}}{A_\pm(n,t)}  \times\\ \label{asympsipm}
& \times
\Big(1 + \Big(B_\pm(n,t) \pm \sum_{j=1}^{n} b_q(j- {\scriptstyle{0 \atop 1}},t) \Big)\frac{1}{z}
+ O(\frac{1}{z^2}) \Big),
\end{align}
as $z \to \infty$, where
\be \label{defABpm}
\aligned
A_+(n,t) &= \prod_{j=n}^{\infty} \frac{a(j,t)}{a_q(j,t)}, \quad
B_+(n,t)= \sum_{j=n+1}^\infty (b_q(j,t)-b(j,t)), \\
A_-(n,t) &= \!\!\prod_{j=- \infty}^{n-1}\! \frac{a(j,t)}{a_q(j,t)}, \quad
B_-(n,t) = \sum_{j=-\infty}^{n-1} (b_q(j,t) - b(j,t)).
\endaligned
\ee

One has the scattering relations
\be \label{relscat}
T(z) \psi_\mp(z,n,t) =  \ol{\psi_\pm(z,n,t)} +
R_\pm(z) \psi_\pm(z,n,t),  \qquad z \in\sigma(H_q),
\ee
where $T(z)$, $R_\pm(z)$ are the transmission respectively reflection coefficients.
Here $\psi_\pm(z,n,t)$ is defined such that 
$\psi_\pm(z,n,t)= \lim_{\eps\downarrow 0}\psi_\pm(z + \I\eps,n,t)$,
$z\in\sigma(H_q)$. If we take the limit from the other side we
have $\ol{\psi_\pm(z,n,t)}= \lim_{\eps\downarrow 0}\psi_\pm(z - \I\eps,n,t)$.

The transmission $T(z)$ and reflection $R_\pm(z)$ coefficients satisfy
\be \label{reltrpm} 
T(z) \ol{R_+(z)} + \ol{T(z)} R_-(z)=0, \qquad |T(z)|^2 + |R_\pm(z)|^2=1.
\ee
In particular one reflection coefficient, say $R(z)=R_+(z)$, suffices.

We will define a Riemann--Hilbert problem on the Riemann
surface $\M$ as follows:
\be
m(p,n,t)= \left\{\begin{array}{c@{\quad}l}
\begin{pmatrix} T(z) \psi_-(z,n,t)  & \psi_+(z,n,t) \end{pmatrix},
& p=(z,+)\\
\begin{pmatrix} \psi_+(z,n,t) & T(z) \psi_-(z,n,t) \end{pmatrix}, 
& p=(z,-)
\end{array}\right..
\ee
Note that $m(p,n,t)$ inherits the poles at $\hat{\mu}_j(0,0)$ and the
essential singularity at $\infty_\pm$ from the Baker--Akhiezer function.

We are interested in the jump condition of $m(p,n,t)$ on $\Sigma$,
the boundary of $\Pi_\pm$ (oriented counterclockwise when viewed from top sheet $\Pi_+$).
It consists of two copies $\Sigma_\pm$ of $\sigma(H_q)$ which correspond to
non-tangential limits from $p=(z,+)$ with $\pm\im(z)>0$, respectively to non-tangential
limits from $p=(z,-)$ with $\mp\im(z)>0$.

To formulate our jump condition we use the following convention:
When representing functions on $\Sigma$, the lower subscript denotes
the non-tangential limit from $\Pi_+$ or $\Pi_-$, respectively,
\be
m_\pm(p_0) = \lim_{ \Pi_\pm \ni p\to p_0} m(p), \qquad p_0\in\Sigma.
\ee
Using the notation above implicitly assumes that these limits exist in the sense that
$m(p)$ extends to a continuous function on the boundary away from
the band edges.

Moreover, we will also use symmetries with respect to the
the sheet exchange map
\be
p^*= \begin{cases}
(z,\mp) & \text{ for } p=(z,\pm),\\
\infty_\mp & \text{ for } p=\infty_\pm,
\end{cases}
\ee
and complex conjugation
\be
\ol{p} = \begin{cases}
(\ol{z},\pm) & \text{ for } p=(z,\pm)\not\in \Sigma,\\
(z,\mp) & \text{ for } p=(z,\pm)\in \Sigma,\\
\infty_\pm & \text{ for } p=\infty_\pm.
\end{cases}
\ee
In particular, we have $\ol{p}=p^*$ for $p\in\Sigma$.

Note that we have $\ti{m}_\pm(p)=m_\mp(p^*)$ for $\ti{m}(p)= m(p^*)$
(since $*$ reverses the orientation of $\Sigma$) and $\ti{m}_\pm(p)= \ol{m_\pm(p^*)}$ for
$\ti{m}(p)=\ol{m(\ol{p})}$.

With this notation, using \eqref{relscat} and \eqref{reltrpm}, we obtain
\be
m_+(p,n,t) = m_-(p,n,t)
\begin{pmatrix} 1-|R(p)|^2 & -\ol{R(p)} \\ R(p) & 1 \end{pmatrix},
\ee
where we have extended our definition of $R$ to $\Sigma$ such that
it is equal to $R(z)$ on $\Sigma_+$ and equal to $\ol{R(z)}$ on $\Sigma_-$.
In particular, the condition on $\Sigma_+$ is just the
complex conjugate of the one on $\Sigma_-$ since we have $R(p^*)= \ol{R(p)}$
and $m_\pm(p^*,n,t)= \ol{m_\pm(p,n,t)}$ for $p\in\Sigma$.

To remove the essential singularity at $\infty_\pm$ and to get a meromorphic
Riemann--Hilbert problem we set
\be \label{defm2}
m^2(p,n,t)=  m(p,n,t)
\begin{pmatrix} \psi_q(p^*,n,t)^{-1} & 0 \\ 0 & \psi_q(p,n,t)^{-1} \end{pmatrix}.
\ee
Its divisor satisfies
\be
(m^2_1) \ge -\dimus{n,t}, \qquad (m^2_2) \ge -\dimu{n,t},
\ee
and the jump conditions become
\begin{align} \nn
m^2_+(p,n,t) &= m^2_-(p,n,t) J^2(p,n,t)\\ \label{rhpm2.1}
J^2(p,n,t) &=
\begin{pmatrix}
1 -|R(p)|^2  &- \ol{R(p) \Theta(p,n,t)}  \E^{-t \phi(p)}\\
R(p)  \Theta(p,n,t) \E^{t \phi(p)} & 1
\end{pmatrix},
\end{align}
where
\be\label{defTheta}
\Theta(p,n,t) = \frac{\theta(\ulz(p,n,t))}{\theta(\ulz(p,0,0))}
\frac{\theta(\ulz(p^*,0,0))}{\theta(\ulz(p^*,n,t))}
\ee
and
\be\label{defsp}
\phi(p,\frac{n}{t}) =
2 \int_{E_0}^p \Omega_0 + 2 \frac{n}{t} \int_{E_0}^p \om_{\infty_+\, \infty_-}
\in \I \R
\ee
for $p\in\Sigma$. Note
\[
\frac{\psi_q(p,n,t)}{\psi_q(p^*,n,t)} = \Theta(p,n,t) \E^{t\phi(p)}.
\]
Observe that
\[
m^2(p) = \ol{m^2(\ol{p})}
\]
and
\[
m^2(p^*) = m^2(\ol{p}) \sigI,
\]
which follow directly from the definition \eqref{defm2}. They are related to the symmetries
\[
J^2(p)=\ol{J^2(\ol{p})} \quad\text{and}\quad J^2(p) = \sigI J^2(p^*)^{-1} \sigI.
\]
Now we come to the normalization condition at $\infty_+$. To this end note
\be \label{m2infp}
m(p,n,t) = \begin{pmatrix}
A_+(n,t) (1 - B_+(n-1,t) \frac{1}{z}) &
\frac{1}{A_+(n,t)}(1 + B_+(n,t) \frac{1}{z} )
\end{pmatrix} + O(\frac{1}{z^2}),
\ee
for $p=(z,+)\to\infty_+$, with $A_\pm(n,t)$ and $B_\pm(n,t)$ are defined in \eqref{defABpm}.
The formula near $\infty_-$ follows by flipping the columns. Here we have used
\be
T(z) = A_-(n,t) A_+(n,t) \Big( 1 - \frac{B_+(n,t) + b_q(n,t) - b(n,t) + B_-(n,t)}{z} + O(\frac{1}{z^2})\Big).
\ee
Using the properties of $\psi(p,n,t)$ and $\psi_q(p,n,t)$ one checks that its divisor satisfies
\be
(m_1) \ge -\dimus{n,t}, \qquad (m_2) \ge -\dimu{n,t}.
\ee
Next we show how to normalize the problem at infinity.
The use of the above symmetries is necessary  and it makes essential use of the second sheet of the
Riemann surface (see also the Conclusion of this paper).

\begin{theorem}\label{thmm3}
The function
\be
m^3(p)= \frac{1}{A_+(n,t)} m^2(p,n,t)
\ee
with $m^2(p,n,t)$ defined in \eqref{defm2} is meromorphic away from $\Sigma$ and satisfies:
\begin{align}\nn
& m^3_+(p) = m^3_-(p) J^3(p), \quad p\in\Sigma,\\ \label{rhpm3}
& (m^3_1) \ge -\dimus{n,t}, \quad (m^3_2) \ge -\dimu{n,t},\\ \nn
& m^3(p^*) = m^3(p) \sigI \\
& m^3(\infty_+) = \begin{pmatrix}  1 & * \end{pmatrix},
\end{align}
where the jump is given by
\be\label{rhpm3.1}
J^3(p,n,t) =
\begin{pmatrix}
1 -|R(p)|^2  &- \ol{R(p) \Theta(p,n,t)}  \E^{-t \phi(p)}\\
R(p)  \Theta(p,n,t) \E^{t \phi(p)} & 1
\end{pmatrix}.
\ee
\end{theorem}

Setting $R(z)\equiv 0$ we clearly recover the purely periodic solution, as we should.
Moreover, note
\be \label{asymm3}
m^3(p) =  \begin{pmatrix}  \frac{1}{A_+(n,t)^2} & 1 \end{pmatrix}
 + \begin{pmatrix}  \frac{B_+(n,t)}{A_+(n,t)^2} & - B_+(n-1,t) \end{pmatrix} \frac{1}{z} +
O(\frac{1}{z^2}).
\ee
for $p=(z,-)$ near $\infty_-$.

While existence of a solution follows by construction, uniqueness follows from
Theorem~\ref{thmrhpQ} and Remark~\ref{remrhpQ}.

\begin{theorem} \label{thmuniq}
The solution of the Riemann--Hilbert problem of Theorem~\ref{thmm3} is unique.
\end{theorem}

\section{The stationary phase points and corresponding contour deformations}
\label{secSPP}

The phase in the factorization problem \eqref{rhpm2.1} is $t\, \phi$ where
$\phi$ was defined in \eqref{defsp}. Invoking \eqref{ominfpm} and
\eqref{Om0}, we see that the stationary phase points are given by 
\be
\prod_{j=0}^g (z -\tilde\lambda_j) + \frac{n}{t} \prod_{j=1}^g (z -\lambda_j) =0.
\ee
Due to the normalization of our Abelian differentials, the numbers
$\lam_j$, $1\le j \le g$, are real and different with precisely one lying in each
spectral gap, say $\lam_j$ in the $j$'th gap.
Similarly, $\ti\lam_j$, $0\le j \le g$, are real and different and
$\ti\lam_j$, $1\le j \le g$, sits in the $j$'th gap. However $\ti\lam_0$ can be
anywhere (see \cite[Sect.~13.5]{tjac}).

As a first step let us clarify the dependence of the stationary phase points
on $\frac{n}{t}$.

\begin{lemma}
Denote by $z_j(\eta)$, $0\le j \le g$, the stationary phase points, where
$\eta=\frac{n}{t}$. Set $\lam_0=-\infty$ and $\lam_{g+1}= \infty$, then
\be
\lam_j < z_j(\eta) < \lam_{j+1}
\ee
and there is always at least one stationary phase point in the $j$'th spectral gap.
Moreover, $z_j(\eta)$ is monotone decreasing with
\be
\lim_{\eta\to-\infty} z_j(\eta) = \lam_{j+1} \quad\text{and}\quad
\lim_{\eta\to\infty} z_j(\eta) = \lam_j.
\ee
\end{lemma}

\begin{proof}
Due to the normalization of the Abelian differential $\Omega_0 + \eta \om_{\infty_+\, \infty_-}$
there is at least one stationary phase point in each gap and they are
all different. Furthermore,
\[
z_j' = - \frac{q(z_j)}{\ti{q}'(z_j)+ \eta q'(z_j)} = -\frac{\prod_{k=1}^g (z_j -\lam_k)}{\prod_{k=0,k\ne j}^g z_j-z_k},
\] 
where
\[
\ti{q}(z) =\prod_{k=0}^g (z -\tilde\lam_k), \quad
q(z)= \prod_{k=1}^g (z -\lam_k).
\]
Since the points $\lam_k$ are fixed points of this ordinary first order differential equation (note that the
denominator cannot vanish since the $z_j$'s are always different), the numbers
$z_j$ cannot cross these points. Combining the behavior as $\eta\to\pm\infty$ with the fact that
there must always be at least one of them in each gap, we conclude that $z_j$ must stay between
$\lam_j$ and $\lam_{j+1}$. This also shows $z_j' <0$ and thus $z_j(\eta)$ is monotone decreasing.
\end{proof}

In summary, the lemma tells us that we have the following picture:
As $\frac{n}{t}$ runs from $-\infty$ to $+\infty$ we start with $z_g(\eta)$
moving from $\infty$ towards $E_{2g+1}$ while the others stay in their
spectral gaps until $z_g(\eta)$ has passed the first spectral band.
After this has happened, $z_{g-1}(\eta)$ can leave its gap, while $z_g(\eta)$
remains there, traverses the next spectral band and so on. Until
finally $z_0(\eta)$ traverses the last spectral band and escapes to
$-\infty$.

So, depending on $n/t$ there is at most 
one single stationary phase point belonging to the union of the bands
$\sigma(H_q)$, say $z_j(n/t)$. 
On the Riemann surface, there are two such points $z_j$ and its flipping
image $z^*_j$ which may (depending on $n/t$) lie in $\Sigma$.

There are three possible cases.

\begin{enumerate}
\item
One stationary phase point, say  $z_j$,  belongs to the interior of a band
$[E_{2j}, E_{2j+1}]$  and all other stationary phase points
lie in open gaps.
\item
$z_j =z_j^* = E_j$ for some $j$ and all other stationary phase points
lie in open gaps.
\item
No stationary phase point belongs to $\sigma(H_q)$.
\end{enumerate}

\subsection*{Case (i)}
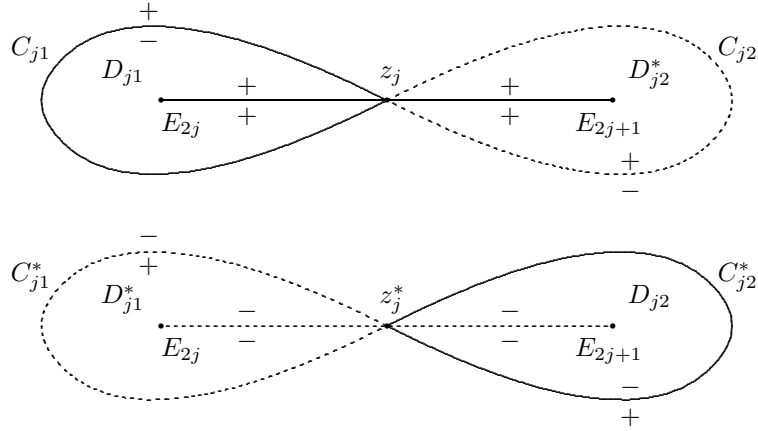
\begin{figure}
\begin{picture}(10,6)

\put(1.2,4.8){$D_{j1}$}
\put(0,5.1){$C_{j1}$}
\put(8.2,4.8){$D_{j2}^*$}
\put(9.4,5.1){$C_{j2}$}
\put(5,4.5){\circle*{0.06}}
\put(4.9,4.8){$z_j$}
\put(2,4.5){\circle*{0.06}}
\put(2,4.1){$E_{2j}$}
\put(8,4.5){\circle*{0.06}}
\put(7.5,4.1){$E_{2j+1}$}

\put(3,4.6){$+$}
\put(3,4.2){$+$}
\put(6.5,4.6){$+$}
\put(6.5,4.2){$+$}

\put(1.7,5.6){$+$}
\put(1.7,5.2){$-$}
\put(8.1,3.6){$+$}
\put(8.1,3.2){$-$}

\put(2,4.5){\line(1,0){6}}
\put(5,4.5){\curve(0, 0, -2.645, -0.951, -4.28, -0.588, -4.28, 0.588, -2.645, 0.951, 0, 0)}
\curvedashes{0.05,0.05}
\put(5,4.5){\curve(0, 0, 2.645, -0.951, 4.28, -0.588, 4.28, 0.588, 2.645, 0.951, 0, 0)}
\curvedashes{}


\put(1.2,1.8){$D_{j1}^*$}
\put(0,2.1){$C_{j1}^*$}
\put(8.2,1.8){$D_{j2}$}
\put(9.4,2.1){$C_{j2}^*$}
\put(5,1.5){\circle*{0.06}}
\put(4.9,1.8){$z_j^*$}
\put(2,1.5){\circle*{0.06}}
\put(2,1.1){$E_{2j}$}
\put(8,1.5){\circle*{0.06}}
\put(7.5,1.1){$E_{2j+1}$}

\put(3,1.6){$-$}
\put(3,1.2){$-$}
\put(6.5,1.6){$-$}
\put(6.5,1.2){$-$}

\put(1.7,2.6){$-$}
\put(1.7,2.2){$+$}
\put(8.1,0.6){$-$}
\put(8.1,0.2){$+$}

\put(5,1.5){\curve(0, 0, 2.645, -0.951, 4.28, -0.588, 4.28, 0.588, 2.645, 0.951, 0, 0)}
\curvedashes{0.05,0.05}
\put(2,1.5){\curve(0,0,6,0)}
\put(5,1.5){\curve(0, 0, -2.645, -0.951, -4.28, -0.588, -4.28, 0.588, -2.645, 0.951, 0, 0)}

\end{picture}
\caption{The lens contour near a band containing a stationary phase point
$z_j$ and its flipping image containing $z_j^*$. Views from the top 
and bottom sheet. Dotted curves lie in the bottom sheet.} \label{fig2}
\end{figure}

Note that in this case
\be\label{phipp}
\phi''(z_j) / \I = \frac{\prod_{k=0,k\ne j}^g (z_j -z_k)}{\I\Rg{z_j}} >0.
\ee
Let us introduce the following "lens" contour
near the band  $[E_{2j}, E_{2j+1}]$ as shown in Figure~\ref{fig2}.
The oriented paths $C_j = C_{j1} \cup C_{j2}$,  $C_j^* = C_{j1}^* \cup C_{j2}^*$
are meant to be close to the band $[E_{2j}, E_{2j+1}]$.

We have
\[
\re(\phi) >0, \quad \text{in } D_{j1},\qquad
\re(\phi) <0, \quad \text{in } D_{j2}.
\]
Indeed
\be
\im(\phi') <0, \quad \text{in } [E_{2j}, z_j], \qquad
\im(\phi') >0, \quad \text{in } [z_{j}, E_{2j+1}]
\ee
noting that $\phi$ is imaginary in $[E_{2j}, E_{2j+1}]$ and writing 
$\phi' = d\phi/dz$. Using the Cauchy-Riemann equations we 
find that the above inequalities are true, as long as 
$C_{j1}, C_{j2}$ are  close enough
to the band $[E_{2j}, E_{2j+1}]$. A similar picture appears in the lower
sheet.

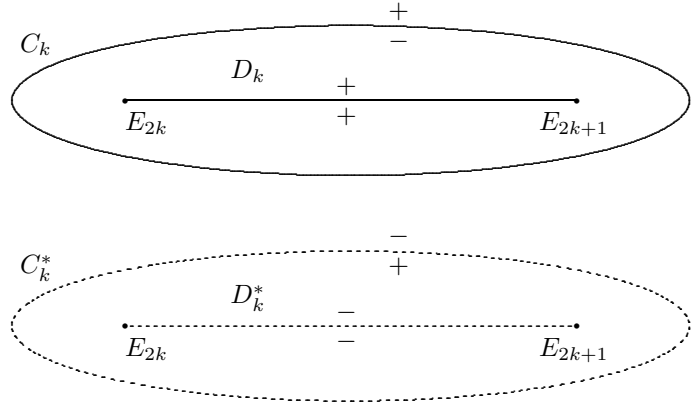
\begin{figure}
\begin{picture}(10,6)

\put(3.4,4.8){$D_k$}
\put(0.6,5.15){$C_k$}
\put(2,4.5){\circle*{0.06}}
\put(2,4.1){$E_{2k}$}
\put(8,4.5){\circle*{0.06}}
\put(7.5,4.1){$E_{2k+1}$}

\put(4.8,4.6){$+$}
\put(4.8,4.2){$+$}

\put(5.5,5.6){$+$}
\put(5.5,5.2){$-$}

\put(2,4.5){\line(1,0){6}}
\put(5,4.5){\curve(0, 1, 2.645, 0.809, 4.28, 0.309, 4.28, -0.309, 2.645, -0.809, 0, -1, %
-2.645, -0.809, -4.28, -0.309, -4.28, 0.309, -2.645, 0.809, 0, 1)}


\put(3.4,1.8){$D_k^*$}
\put(0.6,2.2){$C_k^*$}
\put(2,1.5){\circle*{0.06}}
\put(2,1.1){$E_{2k}$}
\put(8,1.5){\circle*{0.06}}
\put(7.5,1.1){$E_{2k+1}$}

\put(4.8,1.6){$-$}
\put(4.8,1.2){$-$}

\put(5.5,2.6){$-$}
\put(5.5,2.2){$+$}

\curvedashes{0.05,0.05}
\put(2,1.5){\curve(0,0,6,0)}
\put(5,1.5){\curve(0, 1, 2.645, 0.809, 4.28, 0.309, 4.28, -0.309, 2.645, -0.809, 0, -1, %
-2.645, -0.809, -4.28, -0.309, -4.28, 0.309, -2.645, 0.809, 0, 1)}
\end{picture}
\caption{The lens contour near a band not including any stationary phase point.
Views from the top and bottom sheet.} \label{fig3}
\end{figure}

Concerning the  other bands, one simply constructs a "lens" contour
near each  of the other bands $[E_{2k}, E_{2k+1}]$  and $[E_{2k}^*, E_{2k+1}^*]$ 
as shown in Figure~\ref{fig3}.
The oriented paths $C_k, C_k^*$ are meant to be  close
to the band $[E_{2k}, E_{2k+1}]$. The appropriate transformation is now obvious.
Arguing as before, for all bands $[E_{2k}, E_{2k+1}]$ we will have
\[
\re(\phi) < (>) 0, \quad \text{in } D_k, \quad k > (<) j.
\]

Now observe that our jump condition \eqref{rhpm3.1} has the  following important
factorization
\be
J^3=(b_-)^{-1} b_+,
\ee
where 
\[
b_-
=\begin{pmatrix}
1 & \ol{ R  \Theta}  \E^{-t\, \phi} \\ 0 &1
\end{pmatrix}, \qquad
b_+= \begin{pmatrix}
1 & 0 \\ R  \Theta \E^{t\, \phi} & 1
\end{pmatrix}.
\]
This is the right factorization for $z>z_j(n/t)$. Similarly, we have
\be\label{facB}
J^3=(B_-)^{-1} \begin{pmatrix} 1-|R|^2 & 0 \\ 0 & \frac{1}{1-|R|^2}\end{pmatrix} B_+,
\ee
where 
\[
B_-
=\begin{pmatrix}
1 & 0 \\ - \frac{R  \Theta  \E^{t\, \phi}}{1-|R|^2} &1
\end{pmatrix}, \qquad
B_+= \begin{pmatrix}
1 & -\frac{\ol{ R  \Theta}  \E^{-t\, \phi}}{1-|R|^2} \\ 0 & 1
\end{pmatrix}.
\]
This is the right factorization for $z<z_j(n/t)$. To get rid of the diagonal part
we need to solve the corresponding scalar Riemann--Hilbert problem.
Again we have to search for a meromorphic solution. This means
that the poles of the scalar Riemann--Hilbert problem will be added
to the resulting Riemann--Hilbert problem. On the other hand, a
pole structure similar to the one of $m^3$ is crucial for uniqueness.
We will address this problem by choosing the poles of the scalar problem in such
a way that its zeros cancel the poles of $m^3$. The right choice
will turn out to be $\dinuz$ (that is, the Dirichlet divisor corresponding
to the limiting lattice defined in \eqref{limlat}).

\begin{lemma}
Define a divisor $\dinu{n,t}$ of degree $g$ via
\be
\amap(\dinu{n,t}) = \amap(\dimu{n,t}) + \ul{\delta}(n,t),
\ee
where
\be \label{defdel}
\delta_\ell(n,t) = \frac{1}{2\pi\I} \int_{C(n/t)} \log(1-|R|^2) \zeta_\ell.
\ee
Then $\dinu{n,t}$ is nonspecial and $\pi(\hat{\nu}_j(n,t))=\nu_j(n,t)\in\R$ with precisely one
in each spectral gap.
\end{lemma}

\begin{proof}
Using \eqref{ominfpm} one checks that $\delta_\ell$ is real. Hence it follows from \cite[Lem.~9.1]{tjac}
that the $\nu_j$ are real and that there is one in each gap. In particular, the divisor $\dinuz$
is nonspecial by \cite[Lem.~A.20]{tjac}.
\end{proof}

Now we can formulate the scalar Riemann--Hilbert problem required
to eliminate the diagonal part in the factorization \eqref{facB}:
\be \label{rhpd}
\aligned
&d_+(p,n,t) = d_-(p,n,t) (1-|R(p)|^2), \quad p \in C(n/t),\\
&(d)\ge -\dinu{n,t},\\
&d(\infty_+,n,t) = 1,
\endaligned
\ee
where $C(n/t) = \Sigma \cap \pi^{-1}((-\infty,z_j(n/t))$. 
Since the index of the (regularized) jump is zero (see remark below), 
there will be no solution in general unless we admit $g$ additional poles
(see e.g. \cite[Thm.~5.2]{ro}).

\begin{theorem}\label{thmplf}
The unique solution of \eqref{rhpd} is given by
\be \label{defd}
\aligned
d(p,n,t) = & \frac{\theta(\ulz(n,t)+\ul{\delta}(n,t))}{\theta(\ulz(n,t))}
\frac{\theta(\ulz(p,n,t))}{\theta(\ulz(p,n,t)+ \ul{\delta}(n,t))} \times\\
& \times \exp\left( \frac{1}{2\pi\I} \int_{C(n/t)} 
\log (1-|R|^2) \om_{p\, \infty_+}\right),
\endaligned
\ee
where $\ul{\delta}(n,t)$ is defined in \eqref{defdel} and $\om_{p\, q}$
is the Abelian differential of the third kind with poles at $p$ and $q$
(cf.\ Remark~\ref{remabdiff3k} below).

The function $d(p)$ is meromorphic in $\M\setminus\Sigma$ with first order poles at
$\hat{\nu}_j(n,t)$ and first order zeros at $\hat{\mu}_j(n,t)$.
Also $d(p)$ is uniformly bounded in $n,t$ away from the poles.

In addition, we have $d(p)=\ol{d(\ol{p})}$.
\end{theorem}

Note that this formula is different (in fact much simpler) from the explicit solution
formula from Rodin \cite[Sec.~1.8]{ro}. It is the core of our explicit formula
\eqref{limlat} for the limiting lattice.

\begin{proof}
On the Riemann sphere, a scalar Riemann--Hilbert problem is solved by the Plemelj--Sokhotsky formula.
On our Riemann surface we need to replace the Cauchy kernel $\frac{d\lam}{\lam-z}$
by the Abelian differential of the third kind $\om_{p\, \infty_+}$. But now it is
important to observe that this differential
is not single-valued with respect to $p$. In fact, if we move $p$ across the $a_\ell$ cycle,
the normalization $\int_{a_\ell} \om_{p\, \infty_+}=0$ enforces a jump by $2\pi\I\zeta_\ell$.
One way of compensating for these jumps is by adding to $\om_{p\, \infty_+}$
suitable integrals of Abelian differentials of the
second kind (cf.\ \cite[Sec~1.4]{ro} or Appendix~\ref{secSIE}). Since this will produce
essential singularities after taking exponentials we prefer to rather
leave $\om_{p\, \infty_+}$ as it is and
compensate for the jumps (after taking exponentials) by proper use of Riemann theta functions.

To this end recall that the Riemann theta function satisfies
\be\label{thetajump}
\theta( \ulz + \ul{m} + \ul{\tau} \, \ul{n}) = \exp [2 \pi \I \left( -
\spr{\ul{n}}{\ulz} - \frac{\spr{\ul{n}}{\ul{\tau} \, \ul{n}} }{2}\right)]
\theta(\ulz), \quad \ul{n}, \ul{m} \in \Z^g,
\ee
where $\ul{\tau}$ is the matrix of $b$-periods defined in \eqref{deftau} and $\spr{.}{..}$
denotes the scalar product in $\R^g$ (cf., e.g.\ \cite{fk} or \cite[App.~A]{tjac}).
By definition both the theta functions (as functions on $\M$) and the exponential
term are only defined on the "fundamental polygon" $\hat{\M}$ of
$\M$ and do not extend to
single-valued functions on $\M$ in general. However, multi-valuedness apart, $d$
is a (locally) holomorphic solution of our Riemann--Hilbert problem
which is one at $\infty_+$ by our choice of the second pole of the Cauchy kernel
$\om_{p\, \infty_+}$. The ratio of theta functions is, again apart from multi-valuedness,
meromorphic with simple zeros at $\hat{\mu}_j$ and simple poles at $\hat{\nu}_j$
by Riemann's vanishing theorem. Moreover, the normalization is chosen again such
that the ratio of theta functions is one at $\infty_+$. Hence it remains to verify that \eqref{defd}
gives rise to a single-valued function on $\M$.

Let us start by looking at the values from the left/right on the cycle $b_\ell$. Since our
path of integration in $\ul{z}(p)$ is forced to stay in $\hat{\M}$, the difference
between the limits from the right and left is the value of the integral along $a_\ell$.
So by \eqref{thetajump} the limits of the theta functions match. Similarly, since
$\om_{p\, \infty_+}$ is normalized along $a_\ell$ cycles, the limits from the left/right
of $\om_{p\, \infty_+}$ coincide. So the limits of the exponential
terms from different sides of $b_\ell$ match as well.

Next, let us compare the values from the left/right on the cycle $a_\ell$. Since our
path of integration in $\ul{z}(p)$ is forced to stay in $\hat{\M}$, the difference
between the limits from the right and left is the value of the integral along $b_\ell$.
So by \eqref{thetajump} the limits of the theta functions will differ by a multiplicative
factor $\exp(2\pi\I \delta_\ell)$. On the other hand, since $\om_{p\, \infty_+}$ is normalized
along $a_\ell$ cycles, the values from the right and left will differ by $-2\pi\I \zeta_\ell$.
By our definition of $\ul{\delta}$ in \eqref{defdel}, the jumps of the ration of theta
functions and the exponential term compensate each other which shows that
\eqref{defd} is single-valued.

To see uniqueness let $\ti{d}$ be a second solution and consider $\ti{d}/d$. Then
$\ti{d}/d$ has no jump and the Schwarz reflection principle implies that it extends
to a meromorphic function on $\M$. Since the poles of $d$ cancel the poles of $\ti{d}$,
its divisor satisfies $(\ti{d}/d) \ge -\dimuz$. But  $\dimuz$ is nonspecial and thus
$\ti{d}/d$ must be constant by the Riemann--Roch theorem. Setting $p=\infty_+$
we see that this constant is one, that is, $\ti{d}=d$ as claimed.

Finally, $d(p)=\ol{d(\ol{p})}$ follows from uniqueness since both functions solve \eqref{rhpd}.
\end{proof}

\begin{remark}\label{remabdiff3k}
The Abelian differential $\om_{p\, q}$ used in the previous theorem is explicitly given by
\be
\om_{p\, q} = \left( \frac{R_{2g+2}^{1/2} + \Rg{p}}{2(\pi - \pi(p))} -
\frac{R_{2g+2}^{1/2} + \Rg{q}}{2(\pi - \pi(q))} + P_{p q}(\pi) \right)
\frac{d\pi}{R_{2g+2}^{1/2}},
\ee
where $P_{p q}(z)$ is a polynomial of degree $g-1$ which has to be determined from
the normalization $\int_{a_\ell} \om_{p\, p^*}=0$. For $q=\infty_\pm$ we have
\be\label{defompinfpm}
\om_{p\, \infty_\pm} = \left( \frac{R_{2g+2}^{1/2} + \Rg{p}}{2(\pi - \pi(p))} \mp
\frac{1}{2} \pi^g + P_{p \infty_\pm}(\pi) \right)
\frac{d\pi}{R_{2g+2}^{1/2}}.
\ee
\end{remark}

\begin{remark}\label{remd}
Once the last stationary phase point has left the spectrum, that is,
once $C(n/t)=\Sigma$, we have $d(p) = A^{-1} T(z)^{\pm 1}$, $p=(z,\pm)$ (compare \cite{tag}).
Here $A=A_+(n,t) A_-(n,t) = T(\infty)$.
\end{remark}

In particular,
\be \label{defdinfm}
\aligned
d(\infty_-,n,t) = & \frac{\theta(\ulz(n-1,t))}{\theta(\ulz(n,t))}
\frac{\theta(\ulz(n,t)+ \ul{\delta}(n,t))}{\theta(\ulz(n-1,t)+\ul{\delta}(n,t))} \times\\
& \times \exp\left( \frac{1}{2\pi\I} \int_{C(n/t)} 
\log (1-|R|^2) \om_{\infty_-\, \infty_+}\right),
\endaligned
\ee
since $\ulz(\infty_-,n,t)=\ulz(\infty_+,n-1,t)=\ulz(n-1,t)$. Note that $\ol{d(\infty_-,n,t)}=
d(\ol{\infty_-},n,t) = d(\infty_-,n,t)$ shows that $d(\infty_-,n,t)$ is real-valued. Using
\eqref{ominfpm} one can even show that it is positive.

The next lemma characterizes the singularities of $d(p)$ near the stationary phase points
and the band edges.

\begin{lemma} \label{lemd}
For $p$ near a stationary phase point $z_j$ or $z_j^*$ (not equal to a band edge) we have
\be\label{defepm}
d(p)= (z-z_j)^{\pm\I\nu} e^\pm(z), \quad p=(z,\pm),
\ee
where $e^\pm(z)$ is H\"older continuous of any exponent less than $1$ near $z_j$ and
\be
\nu = - \frac{1}{2\pi} \log (1-|R(z_j)|^2)>0.
\ee
Here $(z-z_j)^{\pm\I\nu}=\exp(\pm\I\nu\log(z-z_j))$, where the branch cut of the logarithm is
along the negative real axis.

For $p$ near a band edge $E_k\in C(n/t)$ we have
\be
d(p) = T^{\pm 1}(z) \ti{e}^\pm(z), \quad p=(z,\pm),
\ee
where $\ti{e}^\pm(z)$ is holomorphic near $E_k$ if none of the $\nu_j$ is equal
to $E_k$ and $\ti{e}_\pm(z)$ has a first order pole at $E_k=\nu_j$ else.
\end{lemma}

\begin{proof}
The first claim we first rewrite \eqref{defd} as
\begin{align}\nn
d(p,n,t) = & \exp\left( \I\nu \int_{C(n/t)} \om_{p\,\infty_+}\right) \frac{\theta(\ulz(n,t)+\ul{\delta}(n,t))}{\theta(\ulz(n,t))}
\frac{\theta(\ulz(p,n,t))}{\theta(\ulz(p,n,t)+ \ul{\delta}(n,t))} \times\\ \label{eq:dnearzj}
& \times \exp\left( \frac{1}{2\pi\I} \int_{C(n/t)} 
\log\left(\frac{1-|R|^2}{1-|R(z_j)|^2}\right) \om_{p\,\infty_+}\right).
\end{align}
Next observe
\be
\frac{1}{2} \int_{C(n/t)} \om_{p\, p^*} = \pm\log(z-z_j) \pm \alpha(z_j) + O(z-z_j), \quad p=(z,\pm),
\ee
where $\alpha(z_j)\in\R$, and hence
\be
\int_{C(n/t)} \om_{p\,\infty_+} = \pm\log(z-z_j) \pm \alpha(z_j) + \frac{1}{2} \int_{C(n/t)} \om_{\infty_-\,\infty_+} +O(z-z_j),
\quad p=(z,\pm),
\ee
from which the first claim follows.

For the second claim note that
\[
t(p) = \frac{1}{T(\infty)} \begin{cases} T(z), & p =(z,+)\in\Pi_+,\\ T(z)^{-1}, & p=(z,-)\in\Pi_-,\end{cases}
\]
satisfies the (holomorphic) Riemann--Hilbert problem
\[
\aligned
&t_+(p) = t_-(p) (1-|R(p)|^2), \quad p \in \Sigma,\\
&t(\infty_+) = 1.
\endaligned
\]
Hence $d(p)/t(p)$ has no jump along $C(n,t)$ and is thus holomorphic near $C(n/t)$
away from band edges $E_k=\nu_j$ (where there is a simple pole) by the Schwarz
reflection principle.
\end{proof}

Furthermore,

\begin{lemma}
We have
\be
e^{\pm}(z) = \ol{e^\mp(z)}, \qquad p=(z,\pm)\in\Sigma\backslash C(n/t),
\ee
and
\begin{align}\nn
e^+(z_j) = & \exp\left( \I\nu \alpha(z_j) + \frac{\I\nu}{2} \int_{C(n/t)} \om_{\infty_-\,\infty_+}\right) \times\\ \nn
& \times \frac{\theta(\ulz(n,t)+\ul{\delta}(n,t))}{\theta(\ulz(n,t))}
\frac{\theta(\ulz(z_j,n,t))}{\theta(\ulz(z_j,n,t)+ \ul{\delta}(n,t))} \times\\
& \times \exp\left( \frac{1}{2\pi\I} \int_{C(n/t)} 
\log\left(\frac{1-|R|^2}{1-|R(z_j)|^2}\right) \left(\om_{z_j\, z_j^*} + \om_{\infty_-\,\infty_+}\right)\right),
\end{align}
where
\be\label{def:alzj}
\alpha(z_j) = \lim_{p\to z_j} \frac{1}{2} \int_{C(n/t)} \om_{p\, p^*} - \log(\pi(p)-z_j).
\ee
Here $\alpha(z_j)\in\R$ and $\om_{p\, p^*}$ is real whereas $\om_{\infty_-\,\infty_+}$ is
purely imaginary on $C(n/t)$.
\end{lemma}

\begin{proof}
The first claim follows since $d(p^*) = d(\ol{p}) = \ol{d(p)}$ for $p\in\Sigma\backslash C(n/t)$.
The second claim follows from \eqref{eq:dnearzj} using $\int_{C(n/t)} f\, \om_{p\,\infty_+}
= \frac{1}{2} \int_{C(n/t)} f\, (\om_{p\, p} + \om_{\infty_-\,\infty_+})$ for symmetric functions $f(q)=f(q^*)$.
\end{proof}

Having solved the scalar problem above for $d$ we can introduce
the new Riemann--Hilbert problem
\be \label{defm4}
m^4(p) = d(\infty_-)^{-1} m^3(p) D(p), \quad
D(p) = \begin{pmatrix} d(p^*) & 0 \\
0 &d(p) \end{pmatrix}.
\ee
where $d^*(p)= d(p^*)$ is the unique solution of
\begin{align*}
& d^*_+(p) = d^*_-(p) (1-|R(p)|^2)^{-1}, \quad p \in C(n/t),\\
& (d^*)\ge -\dinus{n,t},\\
& d^*(\infty_-) = 1.
\end{align*}
Note that
\[
\det(D(p)) =  d(p)d(p^*) = d(\infty_-)
\prod_{j=1}^g \frac{z-\mu_j}{z-\nu_j}.
\]
Then a straightforward calculation shows that $m^4$ satisfies
\begin{align}\nn
& m^4_+(p) = m^4_-(p) J^4(p), \quad p\in\Sigma,\\ \label{rhpm4}
& (m^4_1) \ge -\dinus{n,t}, \quad (m^4_2) \ge -\dinu{n,t},\\ \nn
& m^4(p^*) = m^4(p) \sigI \\ \nn
& m^4(\infty_+) = \begin{pmatrix}  1 & * \end{pmatrix},
\end{align}
where the jump is given by
\be
J^4(p) = D_-(p)^{-1} J^3(p) D_+(p), \quad p\in\Sigma.
\ee
In particular, $m^4$ has its poles shifted from $\hat\mu_j(n,t)$ to $\hat\nu_j(n,t)$.

Furthermore, $J^4$ can be factorized as
\be
J^4 = \begin{pmatrix}
1-|R|^2 & - \frac{d}{d^*} \ol{R \Theta} \E^{-t\, \phi}\\
\frac{d^*}{d} R \Theta \E^{t\, \phi} & 1
\end{pmatrix}
= (\ti{b}_-)^{-1} \ti{b}_+, \quad p\in\Sigma\setminus C(n/t),
\ee
where $\ti{b}_\pm = D^{-1} b_\pm D$, that is,
\be\label{defbt}
\ti{b}_- = 
\begin{pmatrix}
1 & \frac{d}{d^*}\ol{R \Theta} \E^{-t\, \phi} \\
0 & 1 \end{pmatrix},\qquad
\ti{b}_+ =
\begin{pmatrix}
1 & 0 \\
\frac{d^*}{d} R \Theta \E^{t\, \phi} & 1 \end{pmatrix},
\ee
for $\pi(p)> z_j(n/t)$ and
\be
J^4 = \begin{pmatrix}
1 & - \frac{d_+}{d^*_-} \ol{R \Theta} \E^{-t\, \phi}\\
\frac{d^*_-}{d_+} R \Theta \E^{t\, \phi} & 1-|R|^2
\end{pmatrix}
= (\ti{B}_-)^{-1} \ti{B}_+, \quad p\in C(n/t),
\ee
where $\ti{B}_\pm = D_\pm^{-1} B_\pm D_\pm$, that is,
\be\label{defBt}
\ti{B}_-  =
\begin{pmatrix} 1 & 0 \\
-\frac{d^*_-}{d_-}  \frac{R  \Theta}{1-|R|^2} \E^{t\, \phi} & 1
\end{pmatrix},\qquad
\ti{B}_+ =
\begin{pmatrix}
1 & - \frac{d_+}{d^*_+} \frac{\ol{R \Theta}}{1-|R|^2} \E^{-t\, \phi} \\
0 & 1 \end{pmatrix},
\ee
for $\pi(p)< z_j(n/t)$.

Note that by $\ol{d(p)}=d(\ol{p})$ we have
\be
\frac{d^*_-(p)}{d_+(p)} = \frac{d^*_-(p)}{d_-(p)} \frac{1}{1-|R(p)|^2} =
\frac{\ol{d_+(p)}}{d_+(p)}, \qquad p\in C(n/t),
\ee
respectively
\be
\frac{d_+(p)}{d^*_-(p)} = \frac{d_+(p)}{d^*_+(p)} \frac{1}{1-|R(p)|^2} =
\frac{\ol{d^*_-(p)}}{d^*_-(p)} , \qquad p\in C(n/t).
\ee

We finally define $m^5$ by
\be \label{defm5}
\aligned
m^5 &= m^4  \ti{B}_+^{-1}, \quad p \in D_k, \: k < j,\\
m^5 &= m^4  \ti{B}_-^{-1}, \quad p \in D_k^*, \: k < j,\\
m^5 &= m^4  \ti{B}_+^{-1}, \quad p \in D_{j1},\\
m^5 &= m^4  \ti{B}_-^{-1}, \quad p \in D_{j1}^*,\\
m^5 &= m^4  \ti{b}_+^{-1}, \quad p \in D_{j2},\\
m^5 &= m^4  \ti{b}_-^{-1}, \quad p \in D_{j2}^*,\\
m^5 &= m^4  \ti{b}_+^{-1}, \quad p \in D_k, \: k > j,\\
m^5 &= m^4  \ti{b}_-^{-1}, \quad p \in D_k^*, \: k > j,\\
m^5 &= m^4, \quad \text{otherwise},
\endaligned
\ee
where we assume that the deformed contour is sufficiently close to
the original one. The new jump matrix is given by
\be \label{rhpm5}
\aligned
m^5_+(p,n,t) &= m^5_-(p,n,t) J^5(p,n,t),\\ 
J^5 &= \ti{B}_+, \quad p \in C_k , ~~k < j,\\
J^5 &= \ti{B}_-^{-1}, \quad p \in C_k^*,~~k < j,\\
J^5 &= \ti{B}_+, \quad p \in C_{j1}, \\ 
J^5 &= \ti{B}_-^{-1}, \quad p \in C_{j1}^{*}, \\
J^5 &= \ti{b}_+, \quad p \in C_{j2}, \\
J^5 &=  \ti{b}_-^{-1}, \quad p \in C_{j2}^*,\\
J^5 &= \ti{b}_+, \quad p \in C_k , ~~k > j,\\
J^5 &= \ti{b}_-^{-1}, \quad p \in C_k^*,~~k > j.
\endaligned
\ee
Here we have assumed that the function $R(p)$ admits an analytic extension in the
corresponding regions. Of course this is not true in general, but we can always evade this
obstacle by approximating $R(p)$ by analytic functions in the spirit of \cite{dz}. We will
provide the details in Section~\ref{secanap}.

The crucial observation now is that the jumps  $J^5$ on the 
oriented paths $C_k$, $C_k^*$ are of the form $\id+ exponentially~small$
asymptotically as $t \to \infty$, at least away from the stationary phase points $z_j$, $z^*_j$.
We thus hope we can simply replace these jumps by the identity matrix (asymptotically as $t \to \infty$)
implying that the solution should asymptotically be given by the constant vector
$\rI$. That this can in fact be done will be shown in the next section by explicitly computing
the contribution of the stationary phase points thereby showing that they are of the order
$O(t^{-1/2})$, that is,
\[
m^5(p) = \rI + O(t^{-1/2})
\]
uniformly for $p$ a way from the jump contour. Hence all which remains to be done to prove
Theorem~\ref{thmMain} and Theorem~\ref{thmMain2} is to trace back the definitions
of  $m^4$ and $m^3$ and comparing with \eqref{asymm3}. First of all, since $m^5$ and $m^4$
coincide near $\infty_-$ we have
\[
m^4(p) = \rI + O(t^{-1/2})
\]
uniformly for $p$ in a neighborhood of $\infty_-$. Consequently, by the definition of $m^4$
from \eqref{defm4}, we have
\[
m^3(p) = d(\infty_-) \begin{pmatrix} d(p^*)^{-1} & d(p)^{-1} \end{pmatrix} + O(t^{-1/2})
\]
again uniformly for $p$ in a neighborhood of $\infty_-$. Finally, comparing this last
identity with \eqref{asymm3} shows
\be
A_+(n,t)^2 = d(\infty_-,n,t) + O(t^{-1/2}), \quad
B_+(n,t) = - d_1(n,t) + O(t^{-1/2}),
\ee
where $d_1$ is defined via
\[
d(p) = 1 + \frac{d_1}{z} + O(\frac{1}{z^2}), \qquad p=(z,+) \text{ near } \infty_+
\]
Hence it remains to compute $d_1$. Proceeding as in \cite[Thm.~9.4]{tjac}
respectively \cite[Sec.~4]{tag} one obtains
\begin{align*}
d_1 =& - \frac{1}{2\pi\I} \int_{C(n/t)} \log(1-|R|^2) \Omega_0\\
& {} - \frac{1}{2}\frac{d}{ds} \log\left( \frac{\theta(\ulz(n,s) + \ul{\delta}(n,t) )}{\theta(\ulz(n,s))} \right) \Big|_{s=t},
\end{align*}
where $\Omega_0$ is the Abelian differential of the second kind defined in \eqref{Om0}.

\subsection*{Case (ii)}

In the  special case where the two stationary phase points coincide
(so $z_j = z_j^*= E_k$ for some $k$) the 
Riemann--Hilbert problem arising above is of a different nature, even in the simpler
non-generic case $|R(E_k)|<1$.
In analogy to the case of the free lattice one expects different local
asymptotics expressed in terms of  Painlev\'e functions.
In the case $|R(E_k)| <1$ the two crosses coalesce and the discussion of Section~\ref{secETFP}
goes through virtually unaltered.
If $|R(E_k)| =1$ the problem is 
singular in an essential way 
and we expect an extra "collisionless shock" phenomenon (on top of the Painlev\'e 
phenomenon) in the region where
$z_j(n/t) \sim E_k$, similar to the one studied
in  \cite{as}, \cite{dvz}, \cite{km2}. The main difficulty
arises from the singularity of  ${R  \over { 1-|R|^2}}$.
An appropriate  "local" Riemann--Hilbert problem however is still 
explicitly solvable and the
actual contribution of the band edges is similar 
to the free case. All this can be studied as in  Section~\ref{secCROSS}
(see also our discussion of this in the Introduction).
But in the present work, we will assume
that the stationary phase points stay away from the $E_k$.

\subsection*{Case (iii)}

In the case where no stationary phase points lie in the spectrum the situation is
similar to the case (i). In fact, it is much simpler since there is no contribution
from the stationary phase points: There is a gap (the $j$-th gap, say) in which two stationary
phase points exist. We construct  "lens-type" contours $C_k$ around every single
band lying to the left of the $j$-th gap and make use of the factorization
$J^3=(\ti{b}_-)^{-1} \ti{b}_+$. We also construct  "lens-type" contours $C_k$ around every single
band lying to the right of the $j$-th gap and make use of the factorization
$J^3=(\ti{B}_-)^{-1} \ti{B}_+$. Indeed, in place of \eqref{defm5}
we set
\be 
\aligned
m^5 &= m^4  \ti{B}_+^{-1}, \quad p \in D_k, \: k < j,\\
m^5 &= m^4  \ti{B}_-^{-1}, \quad p \in D_k^*, \: k < j,\\
m^5 &= m^4  \ti{b}_+^{-1}, \quad p \in D_k, \: k > j,\\
m^5 &= m^4  \ti{b}_-^{-1}, \quad p \in D_k^*, \: k > j,\\
m^5 &= m^4, \quad \text{otherwise}.
\endaligned
\ee

It is now easy to check that in both cases (i) and (iii) 
formula  \eqref{defdinfm} is still true.

\begin{remark}
We have asymptotically reduced our Riemann--Hilbert problem
to one defined on two small crosses. If we are only interested in showing
that the contribution of these crosses is small (i.e that  the solution
of the Riemann--Hilbert problem
is uniformly small for large times) we can evoke the existence theorem in the second
appendix as well as some rescaling argument.

Since we are interested in actually computing the higher order asymptotic term,
a more detailed analysis of the local parametrix Riemann--Hilbert problem
is required.
\end{remark}

\section{The "local" Riemann--Hilbert problems on the small crosses}
\label{secCROSS}

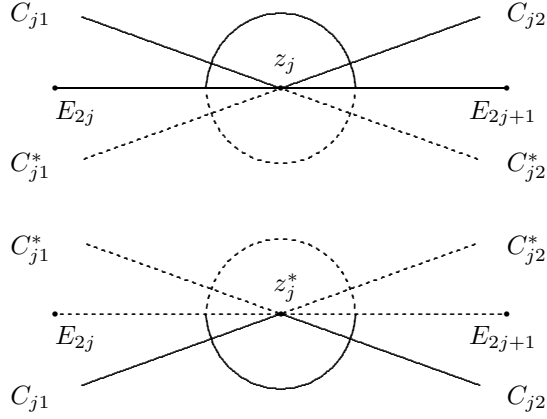
\begin{figure}
\begin{picture}(10,6)

\put(1.4,5.4){$C_{j1}$}
\put(8.0,5.4){$C_{j2}$} 
\put(1.4,3.4){$C_{j1}^*$} 
\put(8.0,3.4){$C_{j2}^*$}  
\put(5,4.5){\circle*{0.06}}
\put(4.9,4.8){$z_j$}
\put(2,4.5){\circle*{0.06}}
\put(2,4.1){$E_{2j}$}
\put(8,4.5){\circle*{0.06}}
\put(7.5,4.1){$E_{2j+1}$}

\put(2,4.5){\line(1,0){6}}
\put(5,4.5){\curve(-2.645, 0.951, 0, 0)}
\put(5,4.5){\curve(2.645, 0.951, 0, 0)}
\put(5,4.5){\curve(1, 0, 0.7, 0.7, 0, 1, -0.7, 0.7, -1, 0)}
\curvedashes{0.05,0.05}
\put(5,4.5){\curve(0, 0, -2.645, -0.951)}
\put(5,4.5){\curve(0, 0, 2.645, -0.951)}
\put(5,4.5){\curve(1, 0, 0.7, -0.7, 0, -1, -0.7, -0.7, -1, 0)}
\curvedashes{}


\put(1.4,2.3){$C_{j1}^*$} 
\put(8.0,2.3){$C_{j2}^*$} 
\put(1.4,0.3){$C_{j1}$} 
\put(8.0,0.3){$C_{j2}$} 
\put(5,1.5){\circle*{0.06}}
\put(4.9,1.8){$z_j^*$}
\put(2,1.5){\circle*{0.06}}
\put(2,1.1){$E_{2j}$}
\put(8,1.5){\circle*{0.06}}
\put(7.5,1.1){$E_{2j+1}$}

\put(5,1.5){\curve(0, 0, 2.645, -0.951)}
\put(5,1.5){\curve(0, 0, -2.645, -0.951)}
\put(5,1.5){\curve(1, 0, 0.7, -0.7, 0, -1, -0.7, -0.7, -1, 0)}
\curvedashes{0.05,0.05}
\put(2,1.5){\curve(0,0,6,0)}
\put(5,1.5){\curve(1, 0, 0.7, 0.7, 0, 1, -0.7, 0.7, -1, 0)}
\put(5,1.5){\curve(2.645, 0.951, 0, 0)}
\put(5,1.5){\curve(-2.645, 0.951, 0, 0)}

\end{picture}
\caption{The small cross containing the stationary phase point
$z_j$ and its flipping image containing $z_j^*$. Views from the top 
and bottom sheet. Dotted curves lie in the bottom sheet.} \label{fig4}
\end{figure}

In the previous section we have shown how the long-time asymptotics can be read off from
the Riemann--Hilbert problem
\begin{align}\nn
& m^5_+(p,n,t) = m^5_-(p,n,t) J^5(p,n,t), \quad p\in\Sigma^5,\\ \nn
& (m^5_1) \ge -\dinus{n,t}, \quad (m^5_2) \ge -\dinu{n,t},\\ \nn
& m^5(p^*,n,t) = m^5(p,n,t) \sigI \\
& m^5(\infty_+,n,t) = \begin{pmatrix}  1 & * \end{pmatrix}.
\end{align}
In this section we are interested in the actual
asymptotic rate at which  $m^5(p) \to \begin{pmatrix}  1 & 1 \end{pmatrix}$.
We have already seen in the previous section that the jumps $J^5$ on the 
oriented paths $C_k$, $C_k^*$ for $k\ne j$ are of the form $\id+ exponentially~small$
asymptotically as $t \to \infty$. The same is true for the oriented paths
$C_{j1}, C_{j2}, C_{j1}^*, C_{j2}^*$ at least away from the stationary phase points $z_j$, $z^*_j$.
On these paths, and in particular near the stationary phase points (see Figure~\ref{fig4}),
the jumps read
{\allowdisplaybreaks \begin{align}\nn
J^5 &= \ti{B}_+ = \begin{pmatrix}
1 & - \frac{d}{d^*} \frac{R^* \Theta^*}{1-R^* R} \E^{-t\, \phi} \\
0 & 1 \end{pmatrix}, \quad p \in C_{j1}, \\ \nn
J^5 &= \ti{B}_-^{-1} = \begin{pmatrix} 1 & 0 \\
\frac{d^*}{d}  \frac{R  \Theta}{1-R^* R} \E^{t\, \phi} & 1
\end{pmatrix}, \quad p \in C_{j1}^{*}, \\ \nn
J^5 &= \ti{b}_+ = \begin{pmatrix} 1 & 0 \\
\frac{d^*}{d} R \Theta \E^{t\, \phi} & 1 \end{pmatrix}, \quad p \in C_{j2}, \\
J^5 &= \ti{b}_-^{-1} = \begin{pmatrix}
1 & -\frac{d}{d^*} R^* \Theta^* \E^{-t\, \phi} \\
0 & 1 \end{pmatrix}, \quad p \in C_{j2}^*.
\end{align}}%
Note that near the stationary phase points the jumps are given by
(cf.\ Lemma~\ref{lemd})
{\allowdisplaybreaks \begin{align}\nn
\hat{B}_+ & = \begin{pmatrix}
1 & - \left(\sqrt{\frac{\phi''(z_j)}{\I}}(z-z_j)\right)^{2\I\nu} \frac{\ol{r}}{1-|r|^2}
\E^{-t\, \phi} \\
0 & 1 \end{pmatrix}, \quad p \in L_{j1}, \\ \nn
\hat{B}_-^{-1} & = \begin{pmatrix} 1 & 0 \\
\left(\sqrt{\frac{\phi''(z_j)}{\I}}(z-z_j)\right)^{-2\I\nu} \frac{r}{1-|r|^2} \E^{t\, \phi} & 1
\end{pmatrix}, \quad p \in L_{j1}^{*}, \\
\hat{b}_+ & = \begin{pmatrix} 1 & 0 \\ \nn
\left(\sqrt{\frac{\phi''(z_j)}{\I}}(z-z_j)\right)^{-2\I\nu}
r \E^{t\, \phi} & 1 \end{pmatrix}, \quad p \in L_{j2}, \\
\hat{b}_-^{-1} &= \begin{pmatrix}
1 & -\left(\sqrt{\frac{\phi''(z_j)}{\I}}(z-z_j)\right)^{2\I\nu}
\ol{r} \E^{-t\, \phi} \\
0 & 1 \end{pmatrix}, \quad p \in L_{j2}^*,
\end{align}}%
where (cf.\ \eqref{defTheta} and \eqref{defepm})
\be\label{defr}
r=R(z_j) \Theta(z_j,n,t) \frac{\ol{e^+(z_j)}}{e^+(z_j)} \left(\frac{\phi''(z_j)}{\I}\right)^{\I\nu}.
\ee
Since the reflection coefficients are continuously differentiable by our decay assumption \eqref{decay}
and by Lemma~\ref{lemd} the error terms will satisfy appropriate H\"older estimates, that is
\be
\| \ti{B}_+(p) - \hat{B}_+(p)\| \le C |z-z_j|^\alpha, \qquad p=(z,+) \in C_{j1},
\ee
for any $\alpha<1$ and similarly for the other matrices.

To reduce our Riemann--Hilbert problem to the one corresponding to the two
crosses we proceed as follows: We take a small disc $D$ around $z_j(n/t)$ and project it
to the complex plane using the canonical projection $\pi$. Now consider the
(holomorphic) matrix Riemann--Hilbert problem {\em in the complex plane} with the very jump
obtained by projection and normalize it to be $\id$ near $\infty$. Denote this solution by
$M(z)$. Then, as is shown in \cite{dz} (see also \cite[Thm.~A.1]{krt2}), the solution of this matrix
Riemann--Hilbert problem on a small cross in the complex plane is asymptotically of the form
\be
M(z) = \id + \frac{M_0}{z-z_j} \frac{1}{t^{1/2}} + O(t^{-\alpha}),
\ee
for any $\alpha<1$ and $z$ outside a neighborhood of $z_j$, where
\begin{align} \nn
M_0 &= \I \sqrt{\I/\phi''(z_j)} \begin{pmatrix} 0 & -\beta(t)\\ \ol{\beta(t)} & 0 \end{pmatrix},\\
\beta(t) &= \sqrt{\nu} \E^{\I(\pi/4-\arg(r)+\arg(\Gamma(\I\nu)))} \E^{-\I t \phi(z_j)} t^{-\I\nu}.
\end{align}
Now we lift this solution back to the small disc on our Riemann-surface by setting
$M(p)=M(z)$ for $p\in D$ and $M(p)=\ol{M(\ol{z})}$ for $p\in D^*$. We define
\be
m^6(p) = \begin{cases}
m^5(p) M^{-1}(p), & p\in D \cup D^*\\
m^5(p), & \text{else}.
\end{cases}
\ee
\\
Note that $m^6$ has no jump inside $D\cup D^*$. Its jumps on the boundary are given by
\be
m^6_+(p) = m^6_-(p) M^{-1}(p), \qquad p\in\partial D \cup \partial D^*
\ee
and the remaining jumps are unchanged. In summary, all jumps outside $D\cup D^*$
are of the form $\id+ exponentially~small$ and the jump on $\partial D \cup \partial D^*$
is of the form $\id + O(t^{-1/2})$. 

In order to identify the leading behaviour it remains to rewrite the Riemann--Hilbert problem for
$m^6$ as a singular integral equation following Appendix~\ref{secSIE}.
Let the operator $C_{w^6}: L^2(\Sigma^6) \to L^2(\Sigma^6) $ be defined by
\be
C_{w^6} f = C_- (f w^6)
\ee
for a vector valued $f$, where $w^6 = J^6 - \id$
and
\be
(C_\pm f)(q) = \lim_{p \to q \in \Sigma^6} \frac{1}{2 \pi \I} \int_{\Sigma^6} f\, \ul{\Omega}_p^\uhnuz, \qquad
\ul{\Omega}_p^\uhnuz = \begin{pmatrix}
\Omega_p^{\uhnuz^*,\infty_+} & 0 \\ 0 & \Omega_p^{\uhnuz,\infty_-}
\end{pmatrix},
\ee
are the Cauchy operators for our Riemann surface. In particular, $\Omega_p^{\uhnuz,q}$
is the Cauchy kernel given by
\be\label{defOm}
\Omega_p^{\uhnuz,q} = \om_{p\, q} + \sum_{j=1}^g I_j^{\uhnuz,q}(p) \zeta_j,
\ee
where
\be
I_j^{\uhnuz,q}(p) = \sum_{\ell=1}^g c_{j\ell}(\uhnuz) \int_q^p \om_{\hnu_\ell,0}.
\ee
Here $\om_{q,0}$ is the (normalized) Abelian differential of the second kind with
a second order pole at $q$ (cf.\ Remark~\ref{remabdiff2k} below).
Note that $I_j^{\uhnuz,q}(p)$ has first order poles at the points $\uhnuz$.

The constants $c_{j\ell}(\uhnuz)$ are chosen such that $\Omega_p^{\uhnuz,q}$ is single
valued, that is,
\be\label{defcjlnu}
\left( c_{\ell k}(\uhnuz) \right)_{1 \le \ell,k \le g} = 
\left( \sum_{j=1}^g c_k(j) \frac{\mu_\ell^{j-1}d\pi}{R_{2g+2}^{1/2}(\hmu_\ell)} \right)_{1 \le \ell,k \le g}^{-1}
\ee
where $c_k(j)$ are defined in \eqref{defcjk} (cf.\ Lemma~\ref{lemck}).

Next, consider the solution $\mu^6$ of the singular integral equation
\be 
\mu = \rI + C_{w^6} \mu  \quad\text{ in }\quad L^2(\Sigma^6).
\ee
Then the solution of our Riemann--Hilbert problem is given by
\be
m^6(p) = 
\rI + \frac{1}{2\pi\I} \int_{\Sigma^6} \mu^6 \, w^6 \, \ul{\Omega}_{p}^\uhnuz.
\ee
Since  $\|w^6\|_\infty = O(t^{-1/2})$ Neumann's formula implies
\be
\mu^6(q) = (\id - C_{w^6})^{-1} \rI = \rI + O(t^{-1/2}).
\ee
Moreover,
\be
w^6(p) = \begin{cases} -\frac{M_0}{z-z_j} \frac{1}{t^{1/2}} + O(t^{-\alpha}), & p \in \partial D,\\
-\frac{\ol{M_0}}{z-z_j} \frac{1}{t^{1/2}} + O(t^{-\alpha}), & p \in \partial D^*.
\end{cases}
\ee
Hence we obtain
\begin{align}\nn
m^6(p) = & \rI - \frac{\rI M_0}{t^{1/2}} \frac{1}{2\pi\I} \int_{\partial D} \frac{1}{\pi-z_j} \, \ul{\Omega}_{p}^\uhnuz\\ \nn
& - \frac{\rI \ol{M_0}}{t^{1/2}} \frac{1}{2\pi\I} \int_{\partial D^*} \frac{1}{\pi-z_j} \, \ul{\Omega}_{p}^\uhnuz
+ O(t^{-\alpha})\\ \nn
= & \rI - \frac{\rI M_0}{t^{1/2}} \ul{\Omega}_{p}^\uhnuz(z_j)
- \frac{\rI \ol{M_0}}{t^{1/2}} \ul{\Omega}_{p}^\uhnuz(z_j^*)
+ O(t^{-\alpha})\\ \nn
= & \rI\\\nn
&- \sqrt{\frac{\I}{\phi''(z_j)t}} \begin{pmatrix} \I \ol{\beta} \Omega_p^{\uhnuz^*,\infty_+}(z_j) - \I \beta\Omega_p^{\uhnuz^*,\infty_+}(z_j^*) &
- \I \beta \Omega_p^{\uhnuz,\infty_-}(z_j) + \I \ol{\beta} \Omega_p^{\uhnuz,\infty_-}(z_j^*) \end{pmatrix}\\
 & + O(t^{-\alpha}).
\end{align}
Note that the right-hand side is real-valued for $p\in\pi^{-1}(\R)\backslash\Sigma$ since $\ol{\Omega_p^{\uhnuz,\infty_\pm}(\ol{q})}
= \Omega_{\ol{p}}^{\uhnuz,\infty_\pm}(q)$ implies
\be
\Omega_p^{\uhnuz,\infty_\pm}(z_j^*) = \ol{\Omega_p^{\uhnuz,\infty_\pm}(z_j)}, \qquad p\in\pi^{-1}(\R)\backslash\Sigma.
\ee
Since we need the asymptotic expansions around $\infty_-$ we note

\begin{lemma}
We have
\be
\Omega_p^{\uhnuz,\infty_+}(z_j)= \Lambda^\uhnuz_0 + \Lambda^\uhnuz_1 \frac{1}{z} + O(\frac{1}{z^2})
\ee
for $p=(z,-)$ near $\infty_-$, where
\be
\Lambda^\uhnuz_0= \Omega_{\infty_-}^{\uhnuz,\infty_+}(z_j) = \Omega_{\infty_-}^{\uhnuz^*,\infty_+}(z_j) =
\om_{\infty_-\, \infty_+}(z_j) + \sum_{k,\ell} c_{k\ell}(\uhnuz) \int_{\infty_+}^{\infty_-} \om_{\hnu_\ell,0} \zeta_k(z_j)
\ee
and
\begin{align}\nn
\Lambda^\uhnuz_1 &= \om_{\infty_-,0}(z_j) +  \sum_{k,\ell} c_{k\ell}(\uhnuz) \om_{\hnu_\ell,0}(\infty_-) \zeta_k(z_j)\\
&= \om_{\infty_-,0}(z_j) -  \sum_{k,\ell} c_{k\ell}(\uhnuz^*) \om_{\hnu_\ell^*,0}(\infty_+) \zeta_k(z_j).
\end{align}
\end{lemma}

\begin{proof}
To see $\Omega_{\infty_-}^\uhnuz(z_j) = \Omega_{\infty_-}^{\uhnuz^*}(z_j)$ note
$c_{k\ell}(\uhnuz^*)= - c_{k\ell}(\uhnuz)$ and
$\int_{\infty_+}^{\infty_-} \om_{\hnu_\ell^*,0}= \int_{\infty_-}^{\infty_+} \om_{\hnu_\ell,0}$.
\end{proof}

Observe that since $c_{k\ell}(\uhnuz)\in\R$ and $\int_{\infty_+}^{\infty_-} \om_{\hnu_\ell,0}\in\R$
we have $\Lambda^\uhnuz_0\in \I\R$.

\begin{remark}\label{remabdiff2k}
Note that the Abelian integral appearing in the previous lemma is explicitly given by
\be
\om_{\infty_-,0} =  \frac{-\pi^{g+1} + \frac{1}{2} \sum_{j=0}^{2g+1} E_j \pi^g + P_{\infty_-,0}(\pi) +
R_{2g+2}^{1/2}}{R_{2g+2}^{1/2}}d\pi,
\ee
with $P_{\infty_-,0}$ a polynomial of degree $g-1$ which has to be determined from the
normalization.

Similarly,
\be
\om_{\hat\nu,0} =  \frac{R_{2g+2}^{1/2}  + R_{2g+2}^{1/2}(\hat\nu) +
\frac{R_{2g+2}'(\hat\nu)}{2R_{2g+2}^{1/2}(\hat\nu)} (\pi-\nu) +
P_{\hat\nu,0}(\pi) \cdot (\pi-\nu)^2 }{2 (\pi-\nu)^2 R_{2g+2}^{1/2}}d\pi,
\ee
with $P_{\hat\nu,0}$ a polynomial of degree $g-1$ which has to be determined from the
normalization.
\end{remark}

As in the previous section, the asymptotics can be read off by using
\be
m^3(p) =  d(\infty_-) m^6(p) \begin{pmatrix} \frac{1}{d(p^*)} & 0 \\ 0 & \frac{1}{d(p)} \end{pmatrix}
\ee
for $p$ near $\infty_-$ and comparing with \eqref{asymm3}.
We obtain
\be
A_+(n,t)^2 = \frac{1}{d(\infty_-)} \left( 1 + \sqrt{\frac{\I}{\phi''(z_j)t}}
\left(\I\ol{\beta} \Lambda^\uhnuz_0 - \I\beta \ol{\Lambda^\uhnuz_0} \right)\right) + O(t^{-\alpha})
\ee
and
\be
B_+(n,t) = - d_1 - \sqrt{\frac{\I}{\phi''(z_j)t}}
\left(\I\ol{\beta} \Lambda^{\uhnuz^*}_1 - \I\beta \ol{\Lambda^{\uhnuz^*}_1} \right) + O(t^{-\alpha}),
\ee
for any $\alpha<1$. Theorem~\ref{thmMain3} and hence also Theorems~\ref{thmMain} and \ref{thmMain2} are
now proved under the assumption that $R(p)$ admits an analytic extension (which will be true if in our decay
assumption \eqref{decay} the weight $n^6$ is replaced by $\exp(-\eps |n|)$ for some $\eps>0$) to be able to make our contour
deformations. We will show how to get rid of this assumption by analytic approximation in the next section.

Summarizing, let us emphasize that the general significance of the method developed in this section is this:
even when a Riemann-Hilbert problem needs to be
considered on an algebraic variety, a localized parametrix Riemann-Hilbert problem
need only be solved in the complex plane and the local solution can then be glued
to the global Riemann-Hilbert solution on the variety. After this gluing procedure the
resulting Riemann-Hilbert problem on the variety is asymptotically small and can 
be solved asymptotically (on the variety) by virtue of the associated singular integral
equations.

The method described in this section can thus provide the higher order asymptotics 
{\em also} in the collisonless shock and Painlev\'e regions mentioned in the Introduction,
by using existing results in (\cite{dvz}, \cite{dz}).

\section{Analytic Approximation}
\label{secanap}

In this section we want to show how to get rid of the analyticity assumption on the reflection coefficient $R(p)$.
To this end we will split $R(p)$ into an analytic part $R_{a,t}$ plus a {\em small} residual term $R_{r,t}$ following
the ideas of \cite{dz} (see also \cite[Sect.~6]{krt2}). The analytic part will be moved to regions of the Riemann surface while the residual
term remains on $\Sigma=\pi^{-1}\big( \sigma(H_q)\big)$. This needs to be done in such a way that the residual term
 is of $O(t^{-1})$ and the growth of the analytic part can be controlled by the decay of the phase.

In order to avoid problems when one of the poles $\nu_j$ hits $\Sigma$, we have to make the
approximation in such a way that the nonanalytic residual term vanishes at the band edges. That is, split $R$ according to
\begin{align} \nn
R(p) =& R(E_{2j}) \frac{z-E_{2j}}{E_{2j+1}-E_{2j}} + R(E_{2j+1}) \frac{z-E_{2j+1}}{E_{2j}-E_{2j+1}}\\ \label{splitR}
& \pm \sqrt{z-E_{2j}} \sqrt{z-E_{2j+1}} \ti{R}(p), \qquad p=(z,\pm),
\end{align}
and approximate $\ti{R}$. Note that if $R \in C^l(\Sigma)$, then $\ti{R} \in C^{l-1}(\Sigma)$.

We will use different splittings for different bands depending on whether the band contains our stationary phase
point $z_j(n/t)$ or not. We will begin with some preparatory lemmas.

For the bands containing no stationary phase points we will use a splitting based on the following Fourier transform 
associated with the background operator $H_q$. Given $R \in C^l(\Sigma)$ we can write
\be
R(p) = \sum_{n\in\Z} \hat{R}(n)\psi_q(p,n,0),
\ee
where $\psi_q(p,x,t)$ denotes the time-dependent Baker--Akhiezer function and (cf.\ \cite{emt}, \cite{emt2})
\be
\hat{R}(n) = \frac{1}{2\pi\I} \oint_\Sigma R(p)  \psi_q(p^*,n,0) \frac{\I \prod_{j=1}^g (\pi(p)-\mu_j)}{\Rg{p}}d\pi(p).
\ee
If we make use of \eqref{defpsiq}, the above expression for $R(p)$ is of the form
\be
R(p)= \sum_{n\in\Z} \hat{R}(n) \theta_q(p,n,0)\exp \big( \I n k(p) \big).
\ee
where $k(p)= -\I \int_{E_0}^p \om_{\infty_+\, \infty_-}$ and $\theta_q(p,n,t)$ collects the remaining
parts in \eqref{defpsiq}.

Using $k(p)$ as a new coordinate and performing $l$ integration by parts one obtains
\be
|\hat{R}(n)| \le \frac{const}{1+|n|^l}
\ee
provided $R \in C^l(\Sigma)$.

\begin{lemma}\label{lem:analapprox}
Suppose $\hat{R} \in \ell^1(\Z)$, $n^l \hat{R}(n) \in \ell^1(\Z)$ and let $\beta>0$ be given.
Then we can split $R(p)$ according to
\[
R(p)= R_{a,t}(p) + R_{r,t}(p),
\]
such that $R_{a,t}(p)$ is analytic for in the region $0 < \im(k(p)) <\eps$ and 
\begin{align}
|R_{a,t}(p) \E^{-\beta t} | &= O(t^{-l}), \quad 0 < \im(k(p)) <\eps,\\
|R_{r,t}(p)| &= O(t^{-l}), \quad p\in\Sigma.
\end{align}
\end{lemma}

\begin{proof}
We choose
\[
R_{a,t}(p) = \sum_{n=-N(t)}^\infty \hat{R}(n) \theta_q(p,n,0)\exp \big( \I n k(p) \big)
\]
with $N(t) = \floor{\frac{\beta_0}{\eps} t}$ for some positive $\beta_0<\beta$. Then, for 
$0< \im(k(p)) <\eps$,
\begin{align*}
\left\vert R_{a,t}(k)\E^{-\beta t} \right\vert 
& \leq C \E^{-\beta t} \sum_{n=-N(t)}^\infty | \hat{R}(n) | \E^{-\im(k(p)) n}\\
& \leq C \E^{-\beta t}\E^{N(t)\eps}\| F \|_1
= \|\hat{R}\|_1 \E^{-(\beta-\beta_0)t},
\end{align*}
which proves the first claim.
Similarly, for $p\in\Sigma$,
\[
\vert R_{r,t}(k) \vert 
\leq C \sum_{n=N(t)+1}^{\infty} \frac{n^l |\hat{R}(-n)|}{n^l}
\leq C \frac{\|n^l \hat{R}(-n)\|_{\ell^1(\N)}}{N(t)^l} 
\leq \frac{\tilde{C}}{t^{l}}  
\]
\end{proof}

For the band which contains $z_j(n/t)$ we need to take the small vicinities of the stationary phase points into account.
Since the phase is cubic near these points, we cannot use it to dominate the exponential growth of the analytic
part away from $\Sigma$. Hence we will take the phase as a new variable and use the Fourier transform
with respect to this new variable. Since this change of coordinates is singular near the stationary phase points,
there is a price we have to pay, namely, requiring additional smoothness for $R(p)$. 

Without loss of generality we will choose the path of integration in our phase $\phi(p)$, defined in \eqref{defsp},
such that $\phi(p)$ is continuous (and thus analytic) in $D_{j,1}$ with continuous limits on the boundary
(cf.\ Figure~\ref{fig2}). We begin with

\begin{lemma}
Suppose $R(p)\in C^5(\Sigma)$. Then we can split $R(p)$ according to
\be
R(p) = R_0(p) +(\pi(p)- \pi(z_j)) H(p), \qquad p \in \Sigma \cap D_{j,1},
\ee
where $R_0(p)$ is a real rational function on $\M$ such that $H(p)$ vanishes  
at $z_j$, $z_j^*$ of order three and has a Fourier series
\be
H(p)=\sum_{n\in\Z} \hat{H}(n) \E^{n \om_0 \phi(p)}, \qquad \om_0= \frac{2\pi\I}{\phi(z_j)-\phi(z_j^*)}>0,
\ee
with $n\hat{H}(n)$ summable. Here $\phi$ denotes the phase defined in \eqref{defsp}.
\end{lemma}

\begin{proof}
We begin by choosing a rational function $R_0(p) = a(z) + b(z) \Rg{p}$ with $p=(z,\pm)$ such that $a(z)$, $b(z)$
are real-valued polynomials which are chosen such that $a(z)$ matches the values of $\re(R(p))$
and its first four derivatives at $z_j$ and $\I^{-1} b(z) \Rg{p}$ matches the values of $\im(R(p))$
and its first four derivatives at $z_j$. Since $R(p)$ is $C^5$ we infer that $H(p)\in C^4(\Sigma)$
and it vanishes together with its first three derivatives at $z_j$, $z_j^*$.

Note that $\phi(p)/\I$, where $\phi$ is defined in \eqref{defsp} has a maximum at $z_j^*$
and a minimum at $z_j$. Thus the phase $\phi(p)/\I$ restricted to $\Sigma \cap D_{j,1}$ gives
a one to one coordinate transform $\Sigma \cap D_{j,1} \to [\phi(z_j^*)/\I, \phi(z_j)/\I]$   
and we can hence express $H(p)$ in this new coordinate. The coordinate  
transform locally looks like a cube root near $z_j$ and $z_j^*$,
however, due to our assumption that $H$ vanishes there, $H$ is still  
$C^2$ in this new coordinate and the Fourier transform
with respect to this new coordinates exists and has the required  
properties.
\end{proof}

Moreover, as in Lemma~\ref{lem:analapprox} we obtain:

\begin{lemma}\label{lem:analapprox2}
Let $H(p)$ be as in the previous lemma. Then we can split $H(p)$ according to
$H(p)= H_{a,t}(p) + H_{r,t}(p)$ such that $H_{a,t}(p)$ is analytic in the region $\re(\phi(p))<0$
and 
\be
|H_{a,t}(p) \E^{\phi(p) t/2} | = O(1), \: p\in \ol{D_{j,1}}, \quad
|H_{r,t}(p)| = O(t^{-1}), \: p\in\Sigma.
\ee
\end{lemma}

\begin{proof}
We choose $H_{a,t}(p) = \sum_{n=-K(t)}^{\infty}\hat{H}(n)\E^{n \om_0 \phi(p)}$ with $K(t) = \floor{t/(2 \om_0)}$.
Then we can proceed as in Lemma~\ref{lem:analapprox}:
\begin{align*}
\vert H_{a,t}(p) \E^{\phi(p) t/2} \vert
\leq \|\hat{H}\|_1 |\E^{-K(t) \om_0 \phi(p)+\phi(p) t/2}|
\leq \|\hat{H}\|_1
\end{align*} 
and 
\[
|H_{r,t}(p)| \leq \frac{1}{K(t)} \sum_{n=K(t)+1}^\infty n |\hat{H}(-n)| \leq \frac{C}{t}.
\]
\end{proof}

Clearly an analogous splitting exists for $p\in\Sigma \cap D_{j2}$.

Now we are ready for our analytic approximation step. First of all recall that our jump is given in terms
$\ti{b}_\pm$ and $\ti{B}_\pm$ defined in \eqref{defbt} and \eqref{defBt}, respectively. While $\ti{b}_\pm$
are already in the correct form for our purpose, this is not true for $\ti{B}_\pm$ since they contain
the non-analytic expression $|T(p)|^2$. To remedy this we will rewrite $\ti{B}_\pm$ in terms of the left
rather than the right scattering data. For this purpose let us use the notation $R_r(p) \equiv R_+(p)$
for the right and $R_l(p) \equiv R_-(p)$ for the left reflection coefficient. Moreover, let
$d_r(p,x,t) = d(p,x,t)$ and $d_l(p,x,t) \equiv T(p)/d(p,x,t)$.

With this notation we have
\be
J^4(p) = \begin{cases}
\ti{b}_-(p)^{-1} \ti{b}_+(p), \qquad \pi(p)> z_j(n/t),\\
\ti{B}_-(p)^{-1} \ti{B}_+(p), \qquad \pi(p)< z_j(n/t),\\
\end{cases}
\ee
where
\begin{align*}
\ti{b}_- &= \begin{pmatrix} 1 & \frac{d_r(p,x,t)}{d_r(p^*,x,t)}R_r(p^*)\Theta(p^*)\E^{-t\phi(p)} \\ 0 & 1 \end{pmatrix}, \\
\ti{b}_+ &= \begin{pmatrix} 1 & 0 \\ \frac{d_r(p^*,x,t)}{d_r(p,x,t)}R_r(p)\Theta(p)\E^{-t\phi(p)}& 1 \end{pmatrix},
\end{align*}
and
\begin{align*}
\ti{B}_- &= \begin{pmatrix} 1 & 0 \\ -\frac{d_{r,-}(p^*,x,t)}{d_{r,-}(p,x,t)}  \frac{R_r(p)  \Theta(p)}{|T(p)|^2} \E^{t\, \phi(p)} & 1 \end{pmatrix}, \\
\ti{B}_+ &= \begin{pmatrix} 1 &  - \frac{d_{r,+}(p,x,t)}{d_{r,+}(p^*,x,t)} \frac{R_r(p^*) \Theta(p^*)}{|T(p)|^2} \E^{-t\, \phi(p)} \\ 0 & 1 \end{pmatrix}.
\end{align*}
Using \eqref{reltrpm} we can write
\begin{align*}
\ti{B}_- &= \begin{pmatrix} 1 & 0 \\ \frac{d_l(p^*,x,t)}{d_l(p,x,t)}R_l(p)\Theta(p)\E^{-t\phi(p)} & 1 \end{pmatrix}, \\
\ti{B}_+ &= \begin{pmatrix} 1 & \frac{d_l(p,x,t)}{d_l(p^*,x,t)}R_l(p^*)\Theta(p^*)\E^{-t\phi(p)} \\ 0 & 1 \end{pmatrix}.
\end{align*}
Now we split $R_r(p)=R_{a,t}(p)+R_{r,t}(p)$ by splitting $\ti{R}_r(p)$ defined via \eqref{splitR} according to Lemma~\ref{lem:analapprox}
for $\pi(p)\in [E_{2k},E_{2k+1}]$ with $k< j$ (i.e., not containing $z_j(n/t)$) and according to Lemma~\ref{lem:analapprox2}
for $\pi(p)\in [E_{2j},z_j(n/t)]$. In the same way we split $R_l(p)=R_{a,t}(p)+R_{r,t}(p)$ for $\pi(p)\in [z_j(n/t),E_{2j+1}]$
and $\pi(p)\in [E_{2k},E_{2k+1}]$ with $k> j$. For $\beta$ in Lemma~\ref{lem:analapprox} we can choose
\be
\beta=\left\{ \begin{array}{ll} \min_{p\in C_k} -\re(\phi(p))>0, & \pi(p)>z_j(n/t),\\
\min_{p\in C_k} \re(\phi(p))>0, & \pi(p)<z_j(n/t). \end{array}\right.
\ee

In this way we obtain
\begin{align*}
\ti{b}_\pm(p) &= \ti{b}_{a,t,\pm}(p) \ti{b}_{r,t,\pm}(p) = \ti{b}_{r,t,\pm}(p) \ti{b}_{a,t,\pm}(p),\\
\ti{B}_\pm(p) &= \ti{B}_{a,t,\pm}(p) \ti{B}_{r,t,\pm}(p) = \ti{B}_{r,t,\pm}(p) \ti{B}_{a,t,\pm}(p).
\end{align*}
Here $\ti{b}_{a,t,\pm}(p)$, $\ti{b}_{r,t,\pm}(p)$ (resp.~$\ti{B}_{a,t,\pm}(p)$, $\ti{B}_{r,t,\pm}(p)$) denote the matrices
obtained from $\ti{b}_\pm(p)$ (resp.~$\ti{B}_\pm(p)$) by replacing $R_r(p)$ (resp.~$R_l(p)$) with $R_{a,t}(p)$, $R_{r,t}(p)$, respectively.
Now we can move the analytic parts into regions of the Riemann surface as in Section~\ref{secSPP}
while leaving the rest on $\Sigma$. Hence, rather than \eqref{rhpm5}, the jump now reads
\be
J^5(p) = \left\{ \begin{array}{ll}
\ti{b}_{a,t,+}(p), & p\in C_k, \quad \pi(p)>z_j(n/t), \\
\ti{b}_{a,t,-}(p)^{-1}, &p\in C_k^*, \quad \pi(p)>z_j(n/t),\\
\ti{b}_{r,t,-}(p)^{-1} \ti{b}_{r,t,+}(p), & p\in\Sigma, \quad \pi(p)>z_j(n/t), \\
\ti{B}_{a,t,+}(p), & p\in C_k, \quad \pi(p)<z_j(n/t), \\
\ti{B}_{a,t,-}(p)^{-1}, &p\in C_k^*, \quad \pi(p)<z_j(n/t),\\
\ti{B}_{r,t,-}(p)^{-1} \ti{B}_{r,t,+}(p), & p\in\Sigma, \quad \pi(p)<z_j(n/t). 
\end{array} \right.
\ee
By construction $R_{a,t}(p) = R_0(p) + (\pi(p)- \pi(z_j)) H_{a,t}(p)$ will satisfy the required
Lipschitz estimate in a vicinity of the stationary phase points (uniformly in $t$) and the
jump will be $J^5(p) = \id+O(t^{-1})$. The remaining parts of $\Sigma$ can be handled analogously
and hence we can proceed as in Section~\ref{secCROSS}.

\section{Conclusion}

We have considered here the stability problem for the periodic Toda lattice under a
short-range perturbation. We have discovered that a nonlinear stationary phase
method (cf.\ \cite{dz}, \cite{km}) is applicable and as a result we have shown that
the long-time behavior of the perturbed lattice is described by a modulated lattice
which undergoes a continuous phase transition (in the Jacobian variety).

We have extended the well-known nonlinear stationary phase method of Deift
and Zhou to Riemann--Hilbert problems living in an algebraic variety.
Even though the studied example involves a hyperelliptic Riemann surface
the  method is easily extended to surfaces with several sheets.
We were forced to tackle such Riemann--Hilbert problems by the very problem,
since there is no way we could use the symmetries needed to normalize the
Riemann--Hilbert problem of Section~\ref{secISTRH} without including a second sheet.
We believe that this 
is one significant  novelty of our contribution.

Although the most celebrated applications of the deformation method
initiated by \cite{dz} for  the asymptotic evaluation of solutions of
Riemann--Hilbert factorization problems have been in the areas
orthogonal polynomials, random matrices and combinatorial probability, 
most mathematical innovations have appeared in the study of nonlinear dispersive
PDEs or systems of ODEs (e.g. \cite{dz}, \cite{dvz}, \cite{kmm}).
It is thus interesting that another mathematical extension of the theory
arises in the study of an innocent looking stability problem for the periodic
Toda lattice.

On the other hand, we see the current work as part of a more general program.
The next step is to consider initial data that are a short pertrubation
of a finite gap solution at $\pm \infty$ but with different genus at each infinity,
a generalized "Toda shock" problem.
Then a similar picture arises (modulation regions separated by "periodic" regions)
but now the genus of the modulated solution can 
also jump between different regions of the $(n,t)$-plane.
The understanding of the more general picture is crucial for the understanding of
the following very interesting problem.

Consider the Toda lattice on the quarter plane $n, t \geq 0$ with 
initial data that are asymptotically
periodic  (or constant) as $n \to \infty$ and periodic  data $a_0(t)$ and $b_0(t)$. What is the long
time behavior of the system?

Special cases of this problem correspond to the generalized Toda shock described above.
A full understanding of the periodic forcing problem thus requires an understanding
of the setting described in this paper.

A related publication is for example \cite{bmk}
where the authors study such a periodic forcing problem 
(for NLS rather than Toda) by extending the 
inverse scattering method of
Fokas (e.g. \cite{fis}) for integrable systems in the quarter plane and actually arrive 
at a Riemann--Hilbert problem living in a Riemann surface.
We thus expect our methods to have a wide applicability.

\appendix

\section{A singular integral equation}
\label{secSIE}

In the complex plane, the solution of a Riemann--Hilbert problem
can be reduced to the solution of a singular integral equation (see \cite{bc})
via a Cauchy-type formula.
In our case the underlying space is a Riemann surface $\mathbb{M}$.
The purpose of this appendix is to produce a more general  Cauchy-type formula  to
Riemann--Hilbert problems of the type
\begin{align}\nn
& m_+(p) = m_-(p) J(p), \quad p\in\Sigma,\\ \label{rhpm}
& (m_1) \ge -\dimuzs, \quad (m_2) \ge -\dimuz,\\ \nn
& m(\infty_+) = m_0\in\C^2.
\end{align}
Once one has such an integral formula, it is easy to "perturb" it and prove that
small changes in the data produce small changes in the solution of the
Riemann-Hilbert problem.

Concerning the jump contour $\Sigma$ and the jump matrix $J$ we will
make the following assumptions:

\begin{hypothesis}\label{hyp:rhp}
Let $\Sigma$ consist of a finite number of smooth oriented finite curves in $\mathbb{M}$
which intersect at most finitely many times with all intersections being transversal.
The divisor $\dimuz$ is nonspecial.
The contour $\Sigma$ does neither contain $\infty_\pm$ nor any of the
points $\uhmuz$ and that the jump matrix $J$ is nonsingular and can be factorized according to
$J = b_-^{-1} b_+ = (\id-w_-)^{-1}(\id+w_+)$, where $w_\pm = \pm(b_\pm-\id)$ are continuous.
\end{hypothesis}

\begin{remark}
(i). We dropped our symmetry requirement 
\be
 m(p^*) = m(p) \sigI
\ee
here since it only is important in the presence of solitons.
However, if both $\Sigma$ and $w_\pm$ are compatible with
this symmetry, then one can restrict all operators below to the corresponding symmetric
subspaces implying a symmetric solution. Details will be given in \cite{krt2}.

(ii). 
The assumption that none of the poles $\uhmuz$ lie on our contour $\Sigma$ can
be made without loss of generality if the jump is analytic since we can move the contour
a little without changing the value at $\infty_-$ (which is the only value we are eventually interested in).
Alternatively, the case where one (or more) of the poles $\hat\mu_j$ lies on $\Sigma$ can be included if one
assumes that $w_\pm$ has a first order zero at $\hat\mu_j$. In fact, in this case one can replace
$\mu(s)$ by $\ti{\mu}(s)=(\pi(s)-\mu_j)\mu(s)$ and $w_\pm(s)$ by $\ti{w}_\pm(s)=(\pi(s)-\mu_j)^{-1}w_\pm(s)$.

Otherwise one could also assume that the matrices $w_\pm$ are H\"older continuous and
vanish at such points. Then one can work with the weighted measure $-\I\Rg{p}d\pi$ on $\Sigma$.
In fact, one can show that the Cauchy operators are still bounded in this weighted Hilbert space
(cf.\ \cite[Thm.~4.1]{gk}).
\end{remark}

Our first step is to replace the classical Cauchy kernel by  a "generalized" Cauchy
kernel  appropriate to our Riemann surface.
In order to get a single valued kernel we need again
to admit $g$ poles. We follow the construction from \cite[Sec.\ 4]{ro}.

\begin{lemma}\label{lemck}
Let $\dimuz$ be nonspecial and introduce the differential
\be\label{defOmpmu}
\Omega_p^\uhmuz = \om_{p\, \infty_+} + \sum_{j=1}^g I_j^\uhmuz(p) \zeta_j,
\ee
where
\be
I_j^\uhmuz(p) = \sum_{\ell=1}^g c_{j\ell}(\uhmuz) \int_{\infty_+}^p \om_{\hmu_\ell,0}.
\ee
Here $\om_{p\,,q}$ is the (normalized) Abelian differential of the third kind with poles
at $p$, $q$ (cf.\ Remark~\ref{remabdiff3k}) and
$\om_{q,0}$ is the (normalized) Abelian differential of the second kind with
a second order pole at $q$ (cf.\ Remark~\ref{remabdiff2k}) and the matrix $c_{j\ell}$
is defined as the inverse matrix of $\eta_\ell(\hmu_j)$, where
$\zeta_\ell = \eta_\ell(z) dz$ is the chart expression in a local chart near $\hmu_j$
(the same chart used to define $\om_{\hmu_j,0}$).

Then $\Omega_p^\uhmuz$ is single valued as a function of $p$ with
first order poles at the points $\uhmuz$.
\end{lemma}

\begin{proof}
Note that $I_j^\uhmuz(p)$ has first order poles at the points $\uhmuz$ hence it
remains to show that the constants $c_{j\ell}(\uhmuz)$ are chosen such that $\Omega_p^\uhmuz$ is single
valued (cf.\ the discussion in the proof of Theorem~\ref{thmplf}). That is,
\[
\int_{b_k} dI_j^\uhmuz = \sum_{\ell=1}^g c_{j\ell} \int_{b_k} \om_{\hmu_\ell,0}
= \sum_{\ell=1}^g c_{j\ell} \eta_k(\hmu_\ell) = \delta_{j k},
\]
where $\zeta_k = \eta_k(z) dz$ is the chart expression in a local chart near
$\hmu_\ell$ (here the $b_k$ periods are evaluated using the usual bilinear relations,
see \cite[Sect.~III.3]{fk} or \cite[Sect.~A.2]{tjac}). That the matrix
$\eta_k(\hmu_\ell)$ is indeed invertible can be seen as follows:
If $\sum_{k=1}^g \eta_k(\hmu_\ell) c_k =0$ for $1\le \ell \le g$, then the divisor of
$\zeta= \sum_{k=1}^g c_k \zeta_k$ satisfies $(\zeta)\ge \dimuz$. But since we
assumed the divisor $\dimuz$ to be nonspecial, $i(\dimuz)=0$, we have
$\zeta=0$ implying $c_k=0$.
\end{proof}

Next we show that the Cauchy kernel introduced in \eqref{defOmpmu} has indeed the correct properties.
We will abbreviate $L^p(\Sigma)=L^p(\Sigma,\C^2)$.

\begin{theorem} \label{thmCrs}
Set
\be
\ul{\Omega}_p^\uhmuz = \begin{pmatrix}
\Omega_p^{\uhmuz^*} & 0 \\ 0 & \Omega_p^\uhmuz
\end{pmatrix}
\ee
and define the matrix operators as follows. Given a $2\times2$ matrix $f$ defined 
on $\Sigma$ with H\"older continuous entries, let
\be
(C f)(p) = \frac{1}{2 \pi \I} \int_{\Sigma} f\, \ul{\Omega}_p^\uhmuz,
\quad\text{for}\quad p \not \in \Sigma,
\ee
and
\be \label{defCpm}
(C_\pm f)(q) = \lim_{p \to q \in \Sigma} (C f)(p)
\ee
from the left and right of $\Sigma$ respectively (with respect to its orientation).
Then 
\begin{enumerate}
\item
The operators $C_\pm$ are given by the Plemelj formulas
\[
\aligned
(C_+ f)(q) - (C_- f)(q) &= f(q),\\
(C_+ f)(q) + (C_- f)(q) &= \frac{1}{\pi \I}\; \dashint_\Sigma f\, \ul{\Omega}_q^\uhmuz,
\endaligned
\]
and extend to bounded operators on $L^2(\Sigma)$. Here $\dashint$ denotes the
principal value integral, as usual.
\item
$C f$ is a meromorphic function off $\Sigma$, with divisor given
by $((C f)_{j1}) \ge -\dimuzs$ and $((C f)_{j2}) \ge -\dimuz$.
\item
$(C f) (\infty_+) =0$.
\end{enumerate}
\end{theorem}

\begin{proof}
In a chart $z=z(p)$ near $q_0\in\Sigma$, the differential $\Omega_q^\uhmuz = 
(\frac{1}{z-z(q)} + O(1)) dz$ and hence the first part follows as in the Cauchy case
on the complex plane (cf.\ \cite{mu} or \cite{stein}) using a partition of unity. To see (ii) note that
the integral over $\om_{p\, \infty_+}$ is a (multivalued) holomorphic function, while
the integral over the rest is a linear combination of the (multivalued) meromorphic
functions $I_j^\uhmuz$ respectively $I_j^{\uhmuz^*}$. By construction, $I_j^\uhmuz$
has at most simple poles at the points $\uhmuz$ and thus (ii) follows.
Finally, to see (iii) observe that $\om_{p\, \infty_+}$ restricted to $\Sigma$ converges
uniformly to zero as $p\to\infty_+$ (cf.\ \eqref{defompinfpm}). Moreover, 
$I_j^{\uhmuz^*}(\infty_+)=0$ and hence (iii) holds.
\end{proof}

Now, let the operator $C_w: L^2(\Sigma) \to L^2(\Sigma) $ be defined by
\be\label{defcw}
C_w f = C_+ (fw_-) + C_- (fw_+)
\ee
for a $2\times2$ matrix valued $f$, where
\[
w_+ = b_+ - \id  \quad\text{ and }\quad   w_- = \id - b_-.
\]

\begin{theorem}\label{thmQ}
Assume Hypothesis~\ref{hyp:rhp} and let $m_0\in\C^2$ be given.

Assume that  $\mu$ solves the singular integral equation
\be \label{musie}
\mu = m_0 + C_w \mu  \quad\text{ in }\quad L^2(\Sigma).
\ee
Then $m$ be defined by the integral formula
\be \label{fQ}
m = m_0 + C (\mu w) \quad\text{ on } \M \setminus \Sigma,
\ee
where $w=w_+ + w_-$, is a solution of the
meromorphic Riemann--Hilbert problem \eqref{rhpm}.

Conversely, if $m$ is a solution of \eqref{rhpm}, then $\mu$
defined via $\mu = m_\pm b_\pm^{-1}$ solves \eqref{musie}.
\end{theorem}

\begin{proof}
Suppose $\mu$ solves \eqref{musie}.
To show that $m$ defined above solves \eqref{rhpm} note that
\[
m_\pm = \id + C_\pm(\mu w).
\]
Thus, using $C_+-C_- = \id$ and the definition of $C_w$ we obtain
\[
\aligned
m_+ &= (m_0 + C_+ (\mu w)) = (m_0+C_+(\mu w_+) + C_+(\mu w_-))\\
& = (m_0+\mu w_+ + C_- (\mu w_+) + C_+ (\mu w_-) )
= (m_0+ \mu w_+ + C_w \mu )\\
& = \mu (\id + w_+) 
\endaligned
\]
and similarly $m_- = \mu (\id - w_-)$. Hence $m_+ b_+^{-1} = \mu = m_- b_-^{-1}$
and thus $m_+ = m_- (b_-)^{-1} b_+$. This proves the jump condition.
That $m$ has the right devisor and the correct normalization at $\infty_+$
follows from Theorem~\ref{thmCrs} (ii) and (iii), respectively.

Conversely, if $m$ is a solution of the Riemann--Hilbert problem \eqref{rhpm},
then we can set $\mu= m_+ b_+^{-1} = m_- b_-^{-1}$ and define $\ti{m}$ by \eqref{fQ}.
To see that in fact $m=\ti{m}$ holds, observe that both satisfy the same additive jump
condition $m_+-m_- = \ti{m}_+-\ti{m}_- = \mu w$. Hence the difference $m-\ti{m}$ has no
jump and thus must be meromorphic. Moreover, by the divisor conditions
$(m_1-\ti{m}_1) \ge -\dimuzs$ and $(m_2-\ti{m}_2) \ge -\dimuz$, the Riemann--Roch
theorem implies that $m-\ti{m}$ is constant. By our normalization at $\infty_+$ this constant
must be the zero vector. Thus $m=\ti{m}$ and as before one computes
\[
m_+ = \mu b_+ - \mu + m_0 + C_w \mu,
\]
showing that \eqref{musie} holds.
\end{proof}

\begin{remark}
(i). The theorem stated above does not address uniqueness. This will be done in
Theorem~\ref{thmrhpQ} under an additional symmetry assumption.

(ii). The notation $b_+, b_-$ is meant to make one think of the example
$J^3 =(b_-)^{-1} b_+ $ in Section~\ref{secSPP}, but the theorem above is fairly general.
In particular it also applies to the 
trivial factorizations $J^3 = \id J^3 = J^3 \id$.
\end{remark}

We  are interested in the  formula \eqref{fQ} evaluated at $\infty_-$.
We write it as
\be \label{ifQ}
\aligned
m(\infty_-) &=  (m_0 + C (\mu w)) (\infty_-)\\ & = 
m_0 + \int_{\Sigma} (\id-C_w)^{-1}(m_0)\, w \, \ul{\Omega}_{\infty_-}^\uhmuz
\endaligned
\ee
and we perturb it with respect to $w$ while keeping the contour $\Sigma$ fixed.

Hence we have a formula for the solution of our Riemann--Hilbert problem $m(z)$ in terms of
$(\id - C_w)^{-1} m_0$ and this clearly raises the question of bounded
invertibility of $\id - C_w$. This follows from Fredholm theory (cf.\ e.g. \cite{zh}):

\begin{lemma}
Assume Hypothesis~\ref{hyp:rhp}.
Then the operator $\id-C_w$ is Fredholm of index zero,
\be
\ind(\id-C_w) =0.
\ee
\end{lemma}

\begin{proof}
Using the Bishop--Kodama theorem \cite{ko} we can approximate $w_\pm$ by
functions which are analytic in a neighborhood of $\Sigma$ and hence, since
the norm limits of compact operators are compact, we can assume that $w_\pm$
are analytic in a neighborhood of $\Sigma$ without loss of generality.

First of all one can easily check that
\be
(\id-C_w) (\id-C_{-w}) = (\id-C_{-w}) (\id-C_w) = \id- T_w,
\ee
where $T_w(f) = C_-[C_-(f w_+)w_+]$. But $ T_w(f)$ is a compact operator.
Indeed, suppose $f_n \in L^2(\Sigma)$ converges weakly to zero.
We will show that $\|T_w f_n\|_{L^2} \to 0$.

Using the analyticity of $w_+$ in a neighborhood of $\Sigma$
and the definition of $C_-$, we can slightly deform
the contour $\Sigma$ to some contour 
$\Sigma'$ close to $\Sigma$, on the right, and
have, by Cauchy's theorem,
\be
T_w f_n(p) = \frac{1}{2 \pi \I}
\int_{\Sigma'} (C(f_n w_+) w_+) \ul{\Omega}_p^\uhmuz.
\ee
Now clearly $(C(f_n w_+) w_+)(p) \to 0$ as $n \to \infty$.
and since also $|(C(f_n w_+) w_+)(p)| < const\, \|f_n\|_{L^2} \|w_+ \|_{L^\infty}
< const$ we infer $\|T_w f_n\|_{L^2} \to 0$ by virtue of the dominated convergence
theorem.

Hence by \cite[Thm.~1.4.3]{proe} $\id-C_w$ is Fredholm. Moreover,
consider $\ind(\id- \eps C_w)$ for $0 \leq \eps \leq 1$ and recall
that $\ind(\id- \eps C_w)$ is continuous with respect to $\eps$ (\cite[Thm.~1.3.8]{proe}).
Since it is an integer, it has to be constant, that is, $\ind(\id-C_w)= \ind(\id) =0$.
\end{proof}

By the Fredholm alternative, it follows that to show the bounded invertibility of $\id-C_w$
we only need to show that $\ker (\id-C_w) =0$. The latter being equivalent to
unique solvability of the corresponding vanishing Riemann--Hilbert problem.

\begin{corollary}\label{cor:fral}
Assume Hypothesis~\ref{hyp:rhp}.

A unique solution of the Riemann--Hilbert problem \eqref{rhpm} exists if and only if the corresponding
vanishing Riemann--Hilbert problem, where the normalization condition is given by
$m(\infty_+)= \begin{pmatrix} 0 & 0\end{pmatrix}$, has at most one solution.
\end{corollary}

We are interested in comparing two Riemann--Hilbert problems associated with
respective jumps $w_0$ and $w$ with $\|w-w_0\|_\infty$ small,
where
\be
\|w\|_\infty= \|w_+\|_{L^\infty(\Sigma)} + \|w_-\|_{L^\infty(\Sigma)}.
\ee
For such a situation we have the following result:

\begin{theorem}\label{thm:remcontour}
Assume that for some data $w_0^t$ the operator
\be
\id-C_{w_0^t}: L^2(\Sigma) \to L^2(\Sigma)
\ee
has a bounded inverse, where the bound is independent of $t$.

Furthermore, assume $w^t$ satisfies
\be
\|w^t - w_0^t\|_\infty \leq \alpha(t)
\ee
for some function $\alpha(t) \to 0$ as $t\to\infty$. Then
$(\id-C_{w^t})^{-1}: L^2(\Sigma)\to L^2(\Sigma)$ also exists
for sufficiently large $t$ and the associated solutions of
the Riemann--Hilbert problems \eqref{rhpm} only differ by $O(\alpha(t))$.
\end{theorem}

\begin{proof}
Follows easily by the Cauchy-type integral formula proved above,
the boundedness of the Cauchy transform  and the second resolvent identity. 

More precisely,
by the boundedness of the Cauchy transform, one has
\[
\|(C_{w^t} - C_{w_0^t})\| \leq const \|w\|_\infty.
\]
Thus, by the second resolvent identity, we infer that $(\id-C_{w^t})^{-1}$ exists for large $t$ and
\[
\|(\id-C_{w^t})^{-1}-(\id-C_{w_0^t})^{-1}\| = O(\alpha(t)).
\]
The claim now follows,  since this implies 
$\|\mu^t - \mu_0^t\|_{L^2} = O(\alpha(t))$ 
where $\mu_0^t$ is defined in the obvious way as in \eqref{musie}
and thus
$m^t(z) - m_0^t(z) = O(\alpha(t))$ uniformly in $z$ away from $\Sigma$.
 \end{proof}

\section{A uniqueness theorem for factorization problems on a Riemann surface}
 \label{secETFP}

In the case where the underlying spectral curve is the complex plane
it is often useful to have a theorem guaranteeing 
existence of a solution of a Riemann--Hilbert problem
under some symmetry conditions. One such is, for example,
the Schwarz reflection theorem  provided in \cite{zh}.
In this section we state and prove an analogous theorem where the underlying
spectral curve is our hyperelliptic curve with real branch cuts.

For any matrix (or vector) $M$ we denote its adjoint (transpose of complex conjugate)
as $M^*$. Then we have

\begin{theorem} \label{thmrhpQ}
Assume in addition to Hypothesis~\ref{hyp:rhp} assume that $\mu_j\in[E_{2j-1},E_{2j}]$
and that $\Sigma$ is symmetric under sheet exchange plus conjugation
($\Sigma= \ol{\Sigma}^*$) such that  

(i) $J(p^*)=  J(\ol{p})^*$, for $p \in \Sigma \setminus \pi^{-1}(\sig(H_q))$,

(ii) $\re(J(p)) = \frac{1}{2}(J(p)+ J(p)^*)$ is positive definite for $p \in \pi^{-1}(\sig(H_q))$,

(iii) $J$ is analytic in a neighborhood of $\Sigma$.

Then the vector Riemann--Hilbert problem \eqref{rhpm} on $\M$ has always a 
unique solution.
\end{theorem}

Note here that the $+$-side of the contour is mapped to the $-$-side under
sheet exchange. In particular, the theorem holds if $J=\id$, 
that is there is no jump, on $\pi^{-1}(\sig(H_q))$.

\begin{proof}
By Corollary~\ref{cor:fral} it suffices to show that the corresponding vanishing problem
has only the trivial solution.

Our strategy is to apply Cauchy's integral theorem to
\[
m(p) m^*(\ol{p}^*) = m_1(p) \ol{m_1(\ol{p}^*)} + m_2(p) \ol{m_2(\ol{p}^*)}.
\]
To this end we will multiply it by a meromorphic differential $d\Omega$
which has zeros at $\ul{\mu}$ and $\ul{\mu}^*$ and simple poles
at $\infty_\pm$ such that the differential $m(p) m^*(\ol{p}^*) d\Omega(p)$
is holomorphic away from the contour.

Indeed let 
\be
d\Omega= -\I \frac{\prod_{j=1}^g (\pi -\mu_j)}{R_{2g+2}^{1/2}} d\pi
\ee
and note that ${\prod_j(z-\mu_j) \over R_{2g+2}^{1/2}(z)}$ is a Herglotz--Nevanlinna
function. That is, it has positive imaginary part in the upper half-plane (and it is
purely imaginary on $\sig(H_q)$). Hence $ m(p) \ol{m^T(p)}d\Omega(p)$ will be positive
on $\pi^{-1}(\sig(H_q))$.

Consider then the integral
\be
\int_D m(p)  m^*(\ol{p}^*) d\Omega(p),
\ee
where $D$ is a $\ol{*}$-invariant contour consisting of one small loop in every
connected component of $\M\setminus\Sigma$. Clearly the above integral
is zero by Cauchy's residue theorem. We will deform $D$ to a $\ol{*}$-invariant
contour consisting of two parts, one, say $D_+$, wrapping around the part of $\Sigma$
lying on $\Pi_+$ and the $+$ side of $\pi^{-1}(\sig(H_q))$ and the other being
$D_- = \ol{D_+}^*$.

For each component $\Sigma_j $ of $\Sigma \setminus \pi^{-1}(\sig(H_q))$
there are two contributions to the integral on the deformed contour: 
\[
\aligned
\int_{ \Sigma_j} m_+(p)  m^*_-(\ol{p}^*) d\Omega & =
\int_{ \Sigma_j} m_-(p) J(p) m^*_-(\ol{p}^*) d\Omega \quad\text{and}\\
\int_{-\Sigma_j} m_-(p) m^*_+(\ol{p}^*) d\Omega &=
\int_{-\Sigma_j} m_-(p) J^*(\ol{p}^*) m^*_-(\ol{p}^*) d\Omega. 
\endaligned
\]
Because of condition (i) the two integrals cancel each other.

In view of the above and using Cauchy's  theorem, one gets
\begin{align} \nn
0 &= \int_D m(p) m^*(\ol{p}^*) d\Omega\\ \nn
&= \int_{\pi^{-1}(\sig(H_q))} [m_+(p) m_-^*(\ol{p}^*) + m_-(p) m_+^*(\ol{p}^*)] d\Omega\\ \nn
&= \int_{\pi^{-1}(\sig(H_q))}  m_-(p) (J(p) +J^*(\ol{p}^*)) m_-^*(\ol{p}^*) d\Omega.
\end{align}
By condition (ii) it now follows that $m_- = 0$  and hence $m = C(\mu w)$ with $\mu = m_-= 0$
by Theorem~\ref{thmQ} (where we used the trivial  factorization $b_- =\id$ and $b_+=J$).
\end{proof}

\begin{remark} \label{remrhpQ}
The same proof also shows uniqueness for the following
symmetric vector Riemann--Hilbert problem on $\M$
\be
\aligned
m_+(p) = m_-(p) J(p), \quad p \in \Sigma,\\
m(p^*) = m(p) \sigI \\
m(\infty_+) = \begin{pmatrix} 1 & * \end{pmatrix}, \quad
 (m_1) \ge -\dimuzs, \quad (m_2) \ge -\dimuz
\endaligned
\ee
where $J(z)$, $\Sigma$, and $\dimuz$ satisfy the same assumptions as in the
previous theorem. Just note that in this case the symmetry assumption
implies $m(p) m^*(\ol{p}^*) = m_1(p) \ol{m_2(\ol{p})} + m_2(p) \ol{m_1(\ol{p})}$.
\end{remark}

\bigskip

\noindent {\bf Acknowledgments.}
We thank I.\ Egorova, H.\ Kr\"uger and A.\ Mikikits-Leitner for pointing out errors in a previous version of this article.
G.T.\ would like to thank P. Deift for discussions on this topic.

S.K.\ gratefully acknowledges the support of the European Science
Foundation (MISGAM program) and the Austrian Science Fund (FWF) during several visits
to the University of Vienna in 2005--2007.
G.T. gratefully acknowledges the extraordinary hospitality of the Courant Institute of Mathematical Sciences,
where part of this research was done.

\end{document}